\documentclass[11pt]{article}

\usepackage{amsmath,amsthm,latexsym,amssymb,bm,amsfonts,mathbbol}

\newcommand{\wt}[1]{\widetilde{#1}}
\renewcommand{\b}[1]{\overline{#1}}
\newcommand{\smallf}[2]{ {\textstyle \frac{#1}{#2}} }

\newcommand{\B}[0]{\mathcal{B}}
\newcommand{\C}[0]{\mathcal{C}}
\newcommand{\M}[0]{\mathcal{M}}
\newcommand{\Z}[0]{\mathcal{Z}}
\renewcommand{\P}[0]{\mathcal{P}}
\renewcommand{\S}[0]{\mathcal{S}}
\newcommand{\N}[0]{\mathcal{N}}
\newcommand{\BB}[0]{\mathbb{B}}
\newcommand{\RR}[0]{\mathbb{R}}
\newcommand{\ZZ}[0]{\mathbb{Z}}
\newcommand{\NN}[0]{\mathbb{N}}
\renewcommand{\O}[0]{\mathcal{O}}
\newcommand{\tot}[0]{\text{tot}}
\newcommand{\tfin}[0]{\mathrm{fin}}
\newcommand{\tmax}[0]{\mathrm{max}}
\newcommand{\tmin}[0]{\mathrm{min}}
\newcommand{\Min}[0]{\mathsf{Min}}
\newcommand{\Bas}[0]{\mathsf{Bas}}

\newtheorem{theorem}[subsection]{Theorem}
\newtheorem{proposition}[subsection]{Proposition}
\newtheorem{lemma}[subsection]{Lemma}
\newtheorem{definition}[subsection]{Definition}
\newtheorem{corollary}[subsection]{Corollary}
\renewenvironment{proof}[0]{\vspace{1.ex} {\it Proof:} \\*[0.7ex]}%
{\hfill $\blacksquare$ \\[0.5ex]}
\begin{document}

\title{Tree Structures: A Variational Approach to Shannon--Wiener
Information}

\author{Hanno Hammer \thanks{email: H.Hammer@umist.ac.uk} \\
Department of Mathematics, UMIST, \\ PO Box 88,
Manchester M60 1QD, UK}

\maketitle

\begin{abstract}
Entanglement measures based on a logarithmic functional form naturally
emerge in any attempt to quantify the degree of entanglement in the
state of a multipartite quantum system. These measures can be regarded
as generalizations of the classical Shannon-Wiener information of a
probability distribution into the quantum regime. In the present work
we introduce a previously unknown approach to the Shannon-Wiener
information which provides an intuitive interpretation for its
functional form as well as putting all entanglement measures with a
similar structure into a new context: By formalizing the process of
information gaining in a set-theoretical language we arrive at a
mathematical structure which we call ''tree structures'' over a given
set. On each tree structure, a tree function can be defined,
reflecting the degree of splitting and branching in the given tree. We
show in detail that the minimization of the tree function on, possibly
constrained, sets of tree structures renders the functional form of
the Shannon-Wiener information. This finding demonstrates that
entropy-like information measures may themselves be understood as the
result of a minimization process on a more general underlying
mathematical structure, thus providing an entirely new
interpretational framework to entropy-like measures of information and
entanglement. We suggest three natural axioms for defining tree
structures, which turn out to be related to the axioms describing
neighbourhood topologies on a topological space. The same minimization
that renders the functional form of the Shannon-Wiener information
from the tree function then assigns a preferred topology to the
underlying set, hinting at a deep relation between entropy-like
measures and neighbourhood topologies.
\end{abstract}

\tableofcontents


%
%
\section{Introduction} \label{Introduction}

In the past two decades, two well-established disciplines within
Mathematics and Physics, namely Information Theory and Quantum
Physics, have merged into a prospective new field on their own. This
process was instigated by advances in Atomic and Molecular Physics
which opened up the possibility of controlling the behaviour of matter
by means of laser light down to very small length scales -- from the
nano regime, involving mesoscopic systems, even further down to the
control of single atoms and molecules by means of appropriate laser
radiation. What is more, even the possibility of reducing the
irradiating source down to nanoscales has sprung up, and controlled
single-photon sources have been experimentally proved to be
possible. Together with these advances, the possibility of utilizing
nanostructures, or even single molecules, as the fundamental building
blocks for future quantum computers, has arisen. The study of
information processing in such an environment, its limitations as well
as its capabilities to exceed classical computational processes, then
has been coined according to the two main pillars contributing to the
new discipline -- Quantum Information Theory.

However, it turns out that the scope of Quantum Information Theory is
much wider than the range of its possible applications within Quantum
Computation might suggest. Indeed, it has been conclusively
demonstrated that Quantum Information Theory provides the main
conceptual as well as computational foundation to tackle some of the
basic unsolved problems within Quantum Mechanics which have haunted
physicists for decades -- the problem of Quantum Nonlocality
\cite{ZeilingerEABell1999a,ZeilingerEA2001a,ANVAKZZ1999}, its presence
signalled by the violation of appropriate Bell-Inequalities
\cite{Bell,Peres1999a,Peres2000a}, and the nature of Entanglement
between distinct quantum systems. These questions have a strong
overlap with the mystery of how a 'classical world' can emerge from a
universe governed by Quantum Mechanics in which all superpositions of
states are allowed, but still only a tiny subset of these, namely
those which we perceive as 'classical', can be ordinarily observed: A
possible answer has been provided by the notion of Decoherence
\cite{ZurekPointerStates1981a,ZurekSupersel1982a,ZurekPreferred1993a,%
Zurek1991PhTo,KieferZehEA}
 --- the decay of quantum correlations between
systems which are subjected to the inevitable quantum noise from an
environment which is ultimately understood to be the universe as a
whole. Here again, the notion of entanglement emerges as a central
quantity.

One of the basic problems in Quantum Information Theory (QIT) is then
to find measures for the quantification of the degree of nonclassical
correlations, or entanglement, between two physical systems. So far,
entanglement has been shown to be quantifiable in two regimes, called
"finite" and "asymptotic"
\cite{HHH2000,DonaldHorodeckiRudolph2001,NielsenThesis}: The first one
attempts to quantify the amount of entanglement within a single copy
of a quantum state; the second one deals with tensor products of a
large number of identical copies of a given state. Many of the
entanglement measures proposed so far have a close relationship to the
classical Shannon-Wiener information
\cite{Shannon1948a,WienerCybernetics,Shannon1949a,ShannonInformation}
of a probability distribution, which is formally identical (up to a
sign) to the thermodynamical entropy of the same distribution. For
example, the so-called ''uniqueness theorem''
\cite{HHH2000,PopescuRohrlich1997,Vidal2000} states that, under
appropriate conditions, all entanglement measures coincide on pure
bipartite states and are equal to the von Neumann entropy
\cite{vonNeumannEntropy} -- the quantum analogue of the classical
Shannon-Wiener information -- of the corresponding reduced density
operators. It is this recurring fact which strongly hints at the
relevance of entropy-like measures of information, such as the
Shannon-Wiener information, both in the classical and in the quantum
context.

In this work we present a new and rather unexpected approach to the
concept of Shannon-Wiener information: We show that this quantity can
be understood as the result of minimization of a so-called {\it tree
function} on a mathematical structure, called {\it tree structure},
which we define and investigate in this work. We show in detail that
the Shannon-Wiener information, as known up to date, may be obtained
as the minimal value of the tree function, when the condition of
minimization is imposed. This puts the notion of Shannon-Wiener
information, and in turn, all other measures of information and
entanglement which are based upon it, into a whole new context, which
is presented in this text. Although tree-like objects are known in
Information Theory \cite{Baierlein}, Complexity Theory and Discrete
Mathematics, the framework presented here is a new and original cast
of this theme, and differs from previously proposed concepts to such
a degree that a complete and self-contained account of the new
structure is justified; this account is given in the present text.

We now give a brief qualitative overview over the new structure and
its main properties:

\paragraph{What are tree structures:}

A tree structure $\B(X)$ over a given set $X$ is a subset of the power
set $\P X$ of $X$ which is obtained by a continuous splitting of its
nodes $b \in \B(X)$ into smaller and ever smaller subsets; this
splitting is described in terms of partitions of sets. Tree structures
can be defined  over sets of arbitrary cardinality, countable or
non-countable. For infinite sets, the tree structures over $X$ are
fractal-like objects \cite{Falconer}. There are three natural axioms
governing tree structures, which are independent of whether the set
$X$ is finite or infinite. These axioms give rise to preferred
topologies on the underlying set $X$.

See sections \ref{Partitions}, \ref{Axioms}.

\paragraph{How do tree structures arise:}

Tree structures arise in modelling processes of information gaining;
they are designed to capture the operational aspect of this
process. In such a model we assign a natural number to the outcome of
an interaction between a unit that seeks to find a distinct but
unknown element $x_0$ of a set $X$, and a unit that possesses this
information, but renders only information about ''neighbourhoods'' of
the distinct element, as these neighbourhoods zoom more and more into
$x_0$. These ''neighbourhoods'' can be given a topological meaning.

See sections \ref{Motivation}, \ref{TreesAndNeighbourhoodTopology}.

\paragraph{What are the typical structural elements:}

The main structural elements are the "nodes" in the tree; these are
subsets of the underlying set $X$ to which two characteristic numbers,
the total number of elements, and the "degree of splitting" in the
next-level partition, are assigned. Strings of such nodes can be
picturized by "paths" in the tree structure. To every path in a finite
tree structure, a natural number, called the "amount" of the path, can
be assigned, which represents the maximal number of Yes-No-questions
that are necessary to single out the element $x$ for the given path in
the given tree. Of central importance is the sum over the amounts of
all complete paths in the tree; this sum will be called the "amount
function". In this way, every tree structure over a given set $X$ can
be assigned a unique value of the amount function. The amount
functions are related to the "tree function" which is a sum over
cardinal numbers and degrees of splitting at every node in the
tree. On certain subsets of trees, amount functions and tree functions
coincide.

See sections \ref{Paths}, \ref{MinPartition}, \ref{Amountunctions},
\ref{TreeFunction}.

\paragraph{The natural question concerning tree structures:}

Assigning a value of the tree function to every tree over $X$, we can
ask on which trees the tree function takes its minimum. This question
can be generalized, as constraints on the admissible trees can be
imposed. The admissible trees then may be chosen so as to preserve a
prescribed initial partition of $X$, which reflects a choice of
"weights" $(w_i)$ for the path amounts in such a tree. This is
analogous to choosing a probability distribution $(p_i)$ for the paths
in the admissible trees.

See sections \ref{MinimalClasses}, \ref{OptimalTrees},
\ref{MinimalClassesTNM}, \ref{OptimalAmount},
\ref{MinimalityOptimalDivision}, \ref{RestrictedMinimalProblems}.

\paragraph{The first main result concerning tree structures:}

If there are no constraints, then the minimal value of the tree
function is close to $n \cdot \lg(n)$, where $\lg(n)$ is an integer
approximation to the logarithm of $n$ with  respect to the basis
$2$. Thus, the mean value of the amounts of $n$ paths in a complete
tree over $X$ comes close to $\lg(n)$, which is the information gained
in finding a distinct element among $n$ "equally weighted" elements;
or the entropy of $n$ distinct states, depending on the context. One
of the central results of this work is that the functional form
$\lg(n)$ of the entropy so defined is itself the result of a process
of minimization, i.e. there is a more general functional form
underlying, namely the tree function. These results can be generalized
to the constrained case; here the terminal elements $b_i \in \B(X)$
are endowed with weights $w_i$, so that the value of the tree function
contains expressions like $n \cdot \lg(n) - \sum  w_i \cdot
\lg(w_i)$. Here we recognize the Shannon-Wiener information, or
entropy,
\begin{equation*}
-\sum \frac{w_i}{n} \cdot \log_2 \left( \frac{w_i}{n} \right) \simeq
\frac{1}{n} \cdot \left[ n \cdot \lg(n) -\sum w_i \cdot \lg(w_i)
  \right]
\end{equation*}
of the probability distribution $\frac{w_i}{n}$, depending on the
context. Again, we have the striking result that the functional form
of the entropy is itself the result of a process of optimization of a
more general expression, namely the tree function, and the entropy, as
usually known, is only the minimal value of this more general
function.

See sections \ref{BasesIntegerLogarithm}, \ref{OptimalAmount},
\ref{MinimalityOptimalDivision}, \ref{MeanPathQuadDeviation},
\ref{RestrictedMinimalProblems}.

\paragraph{The second main result concerning tree structures:}

Every tree structure over a set $X$ defines a neighbourhood topology
\cite{Brown} on $X$. As we vary the tree structures, so vary the
topologies on $X$. A tree function on a given set of, possibly
constrained, tree structures will single out preferred neighbourhood
topologies, namely  those for which the tree function becomes
minimal. This defines something like an action principle for
neighbourhood topologies on the set $X$, where the value of the action
= tree function on the minimal trees is an entropy-like quantity.

See section \ref{TreesAndNeighbourhoodTopology}.

\vspace{1em}

--- The plan of this report is as follows: In section~\ref{Partitions}
we recall the definition of partitions of sets. In
section~\ref{Motivation} we outline how a tree structure encodes the
operational aspect of the problem of information gaining, which yields
the concept of entropy/information. Three natural axioms describing
tree structures are presented in section~\ref{Axioms}. In
section~\ref{OrderedSets} we recall some elementary facts on ordered
sets; in section~\ref{ZPartiallyOrdered} we show how the set of all
partitions of a given set $X$ is a partially ordered set. The concept
of subtrees is discussed in section~\ref{Subtrees}, while ideas
concerning the sum, union, extension, reduction and completion of
trees are introduced in section~\ref{OperationsOnTrees}. After this
preparation we define paths in a tree structure in
section~\ref{Paths}. In section~\ref{PartitionsCompatible} we show how
a tree over $X$ selects a distinct subset of partitions of the
underlying set $X$; here we introduce the important concepts of
minimal and maximal partitions of the underlying set $X$ in the tree
$\B$, and the number $m(b)$ characterising the degree of splitting of
a node in the tree. Then we come to the central notions in our theory:
In section~\ref{Amountunctions} we introduce amount functions on sets
of tree structures. The technical section~\ref{InducedPartitions}
contains a splitting theorem for amount functions. In
section~\ref{TreeFunction} we define the tree function on the set of
all tree structures over $X$. The problem of minimizing trees is first
taken up in section~\ref{MinimalClasses}. We then introduce the
concept of divisions in section~\ref{Divisions}, and explain its
relation to partitions in section~\ref{PartitionsAndDivisions}. In
sections~\ref{OptimalDivision} and~\ref{OptimalTrees} we introduce
optimal divisions of sets, and the concept of optimal trees based on
optimal divisions. Section~\ref{MinimalClassesTNM} defines minimal
classes based on prescribed
divisions. Section~\ref{BasesIntegerLogarithm} introduces the integer
approximation of the logarithm with respect to the base $2$, together
with some of its properties. The value of the optimal amount of an
optimal tree is derived
in~\ref{OptimalAmount}. Section~\ref{PreoptimizedTrees} introduces the
notion of preoptimized trees, which is a tool of central import to the
proof of the minimality of optimal trees. The latter result is
approached in a series of propositions given in
section~\ref{MinimalityOptimalDivision}. Section~\ref{MeanPathQuadDeviation}
reflects the same statements from the point of view of the mean path
amount in a tree over $X$. In section~\ref{IsomorphicTreeStructures}
we introduce an important notion of structural similarity, encaptured
in an appropriate definition of isomorphism between trees. In
section~\ref{RestrictedMinimalProblems} we outline how to find
constrained minimal trees on which the functional form of the tree
function contains expressions like $-\sum w_i\lg \left(
w_i\right)$. In the last section~\ref{TreesAndNeighbourhoodTopology},
we show how tree structures define neighbourhood topologies on $X$,
and how the tree function selects distinct topologies according to a
minimal principle.

This report is based on the preprint \cite{HammerTreePreprint}.


\paragraph{Notation convention:}
For the difference of two sets, which is commonly denoted as
\begin{equation}
\label{NotConvent0}
 A \backslash B = \big\{ x \in A \big| x \neq B \big\}
\end{equation}
we shall use the notation $A-B \equiv A \backslash B$ instead.

\section{Partitions} \label{Partitions}

Let $X$ be a non-empty set. A {\it partition} $z$ of $X$ is a system
of mutually disjoint non-empty subsets $\mu \subset X$ whose union is
$X$, i.e.
\begin{description}
\item[(P1)]  $\bigcup\limits_{\mu \in z} \mu = X$ ,
\item[(P2)]  $\mu \neq \mu' \; \Rightarrow \; \mu \bigcap \mu' =
\emptyset$ .
\end{description}
The {\it power set} $\P X$ of $X$ is the set of all subsets of $X$,
including the empty set. That is to say, $\P X$ contains the elements
\begin{equation}
\label{power1}
 \emptyset \; , \; \{x\} \text{ for $x \in X$} \; , \; \{x,y\} \text{
for $x \neq y \in X$} \; , \; \ldots \; , \; X \quad.
\end{equation}
We see that every partition is a subset $z \subset \P X$ of the power
set of $X$.

The partition $z_0 \equiv \{X\}$ will be called the {\it trivial
partition}. The partition $z$ is said to be {\it complete} if every
element $\mu $ of $z$ contains precisely one element of $X$,
i.e. $\#\mu = 1$ for all $\mu \in z$, or
\begin{equation}
\label{complete1}
 z = \Big\{\, \{x\} \in \P X \, \Big|\, x \in X\, \Big\} \quad.
\end{equation}

The set of all partitions $z$ of $X$ will be denoted by $\Z(X)$. The
set of all nontrivial partitions will be denoted by $\Z^*(X)$,
i.e. $\Z^*(X) = \big\{\, z \in \Z(X)\, \big|\, z \neq \{X\}\, \big\}$.

\section{Movitation for tree structures} \label{Motivation}

We want to show how tree structures arise in the course of modelling
processes of information gaining. We now describe such a model: Let
$X$ be a non-empty finite set, $0 < n \equiv \#X< \infty$. Let $x_0
\in X$ be arbitrary. We want to find a numerical measure for the
information that is gained when $x_0$ has been identified as a
distinct object amongst $n$ objects. Consider the interaction of two
(information processing) units, the first one (storage unit) of which
has stored the knowledge about $x_0$, and the second unit tries to
identify $x_0$ amongst all $n$ elements of $X$. The only knowledge
permitted to the second unit (search unit) is that all $n$ choices are
equally likely. The search unit starts by suggesting a partition of
$X$ to the storage unit; if the number of elements in the partition is
$m_1$, then the search unit has to pose at most $(m_1-1)$ yes-no
questions to the storage unit in order to identify the element of the
partition that contains $x_0$. Next the search unit suggests a
partition of the subset that contains $x_0$, and so on. This gives the
following scheme: On level $1$, we have a partition $z(X) \in \Z(X)$
with $\#z (X) = m_1$ elements, i.e.
\begin{equation}
\label{motiv1}
 z(X) = \big\{ X_1, \ldots, X_{m_1} \big\} \quad. 
\end{equation}
On level $2$ of the emerging tree we partition all the subsets $X_i$
in (\ref{motiv1}): We decompose $X_1$ into $m_2(1)$ non-empty subsets,
$X_2$ into $m_2(2)$ non-empty subsets, $\ldots$, $X_{m_1}$ into
$m_2(m_1)$ subsets; here the subscripts $1,2$ in $m_1, m_2$ refer to
the levels $1$ and $2$, respectively. Hence for $i_1 = 1, \ldots, m_1$
we have partitions $z(X_{i_1}) \in \Z(X_{i_1})$ with cardinality
$\#z(X_{i_1}) \equiv m_2(i_1)$,
\begin{equation}
\label{motiv2}
 z(X_{i_1}) = \big\{ X_{i_1,1}, \ldots, X_{i_1,m_2(i_1)} \big\} \quad.
\end{equation}
Now we continue along these lines: $X_{1,1}$ is decomposed into
$m_3(1,1)$ subsets; $X_{1,m_2(1)}$ is decomposed into $m_3(1,m_2(1))$
subsets; $\ldots$; $X_{m_1, m_2(m_1)}$ is decomposed into
\begin{equation*}
 m_3(m_1,m_2(m_1))
\end{equation*}
subsets, i.e. for $i_1 = 1, \ldots, m_1$, $i_2 = 1, \ldots, m_2(i_1)$
we introduce a partition $z(X_{i_1,i_2}) \in \Z(X_{i_1,i_2})$ with
cardinal number $\#z(X_{i_1,i_2}) = m_3(i_1,i_2)$ such that
\begin{equation}
\label{motiv3}
 z(X_{i_1,i_2}) = \big\{ X_{i_1,i_2,1}, \ldots,
 X_{i_1,i_2,m_3(i_1,i_2)} \big\} \quad,
\end{equation}
etc. Any of the subsets $X_{i_1,i_2\cdots}$ emerging in this process
is an element of the power set $\P X$ of $X$. The totality of all
these subsets is a certain subset of the power set of $X$ which we
shall term a {\it tree structure} or simply a {\it tree $\B(X)$ over
$X$}. Hence,
\begin{equation}
\label{motiv4}
\begin{aligned}
 \B(X) = \Big\{\, & X \, ,  \\
 & X_1\, , \ldots\, , \, X_{m_1}\, , \\
 & X_{1,1}\, , \ldots\, , X_{1,m_2(1)}\, , \ldots\, , \, X_{m_1,1}\, ,
 \, X_{m_1, m_2(m_1)} \, , \ldots\, \Big\} \quad.
\end{aligned}
\end{equation}
We see that the elements of a given tree structure $\B(X)$ can
obviously be labelled by series of the form
\begin{equation}
\label{motiv5}
\begin{aligned}
 & \emptyset\, , \\
 & \big(1\big) \, , \, \ldots \, , \, \big( m_1 \big)\, , \\
 & \big( 1,1 \big)\, , \, \big( 1,m_2(1) \big)\, , \, \ldots\, , \,
 \big( m_1,1\big)\, ,\, \big(m_1, m_2(m_1)\big)\, , \ldots\, \quad.
\end{aligned}
\end{equation}
If the set $X$ is infinite, there can be series (\ref{motiv5}) which
extend forever. On the other hand, if $X$ is finite, then each of
these series is finite and can be denoted in the form $(i_1, \cdots,
i_\kappa)$. In this case the cardinalities
\begin{equation}
\label{motiv6}
 n(i_1, \ldots, i_{\kappa}) \equiv \# X_{i_1, \ldots, i_{\kappa}}
\end{equation}
are natural numbers.

Let $b$ be a {\it terminal} element in the tree, with series $(i_1,
\ldots, i_{\kappa})$; this series may be called {\it complete} if
$n(i_1, \ldots, i_{\kappa}) = 1$, otherwise it will be called {\it
incomplete}. Hence, in a tree over a finite set $X$ with $n= \# X$ we
can have at most $n$ distinct, complete series.

A tree $\B(X)$ may be called {\it complete} if all series associated
with terminal nodes are complete.

For a given set $X$ let $\M(X)$ denote the set of all tree structures
over $X$. The set of all complete tree structures over $X$ will be
denoted by $\C(X)$. Clearly, $\C(X) \subsetneqq \M(X)$ for $\# X \ge
2$.

\vspace{1.ex}

--- We have seen how tree structures emerge naturally in processes
modelling information gaining. The basic properties of tree
structures, as they present themselves from the above analysis, will
be compiled in the next section.

\section{Axiomatic definition of tree structures} \label{Axioms}

We now suggest three natural axioms defining a tree structure as a set
of subsets of $X$, as motivated in eq.~(\ref{motiv4}): A {\it tree
structure $\B(X)$ over $X$} is a system of non-empty subsets $b
\subset X$ of $X$ (hence a subset of the power set $\P X$ of $X$) such
that the following axioms hold:
\begin{description}
\item[(A1)]\; $X \subset \B(X)$.
\item[(A2)]\; If $b, b' \in \B(X)$, then $b \subset b'$ or $b'
\subsetneqq b$ or $b \cap b' = \emptyset$ \; [This is "exclusive or"].
\item[(A3)]\; For all $b, b' \in \B(X)$ there exists $\wt{b} \in
\B(X)$ such that $b, b' \subset \wt{b}$.
\end{description}

Elements $b \in \B(X)$ will be called the {\it nodes in the tree
$\B(X)$}. An element $b \in \B(X)$ will be called {\it primitive}, if
$b$ contains only one element, i.e., $b= \{y\}$ for some $y \in
X$. The tree structure $\B(X)$ will be called {\it complete} if it
contains all primitive elements, i.e., $\{y\} \in \B(X)$ for all $y
\in X$. This definition is clearly consistent with the notion of
completeness as given in the previous section~\ref{Motivation}.

An element $b \in \B(X)$ will be called {\it refinable} if $b$ is not
primitive; hence there exists $b' \subsetneqq b$. If none of the
subsets $b'$ which refine $b$ lie in the given tree $\B(X)$, we call
the element $b$ {\it terminal in $\B(X)$}; in this case we shall also
use the notation $b= b_{\tfin}$. Thus, each node $b \in \B(X)$ which
is not terminal is refinable in the given tree. On the other hand, all
primitive elements are trivially non-refinable, and hence must be
terminal in $\B(X)$.

Most of the definitions we will introduce in this work will be stated
as general as possible, although our actual conclusions regarding the
Shannon-Wiener information will be worked out on finite sets only.

\section{Ordered sets} \label{OrderedSets}

We recall some general definitions regarding ordered sets:

A non-empty set $X$ is called {\it ordered}, if a relation $"\prec"$
is defined on $X$, satisfying:
\begin{description}
\item[(O1)]\; For any two elements $a,b$ of $X$ either $a \prec b$ or
$b \prec a$ or $a=b$ is true.
\item[(O2)]\; If $a\preceq b$ and $b\preceq c$ then $a\preceq c$.
\end{description}

If the non-empty set $X$ contains a non-empty ordered subset $T$, then
$X$ is said to be {\it partially ordered}. Hence every ordered set is
partially ordered. To distinguish this case from a partial ordering we
sometimes say that an ordered set $X$ is {\it totally ordered}.

If $X$ contains an element $x_0$ for which $x_0 \prec x$ for all $x
\in X$ is true, we call $x_0$ the {\it principal element in $X$} [or
in the pair $(X,\prec)$, to be precise].

\section{$\Z(X)$ as a partially ordered set} \label{ZPartiallyOrdered}

On the set $\Z(X)$ of all partitions of $X$, a natural partial
ordering $"\prec"$ can be introduced as follows: Let $z, z' \in
\Z(X)$. The relation $z \prec z'$ is defined to be true if and only if
every $b' \in z'$ is contained in some $b \in z$ according to $b'
\subset b$, and there exists $b \in z$, $b' \in z'$ for which this
inclusion is proper, $b' \subsetneqq b$. In this case we say that the
partition $z'$ is a {\it refinement} of the partition $z$. If both $z$
and $z'$ are finite this implies in particular that $\#z < \#z'$.

Given two partitions $z, z'$, clearly none of the relations $z \prec
z'$ or $z' \prec z$ or $z' = z$ need to be true; this is why the set
$\Z(X)$ is only partially ordered.

If the partition $z$ of $X$ is kept fixed, we can think of the set of
all partitions $\wt{z}$ of $X$ for which $z$ is a refinement, $\wt{z}
\prec z$; they comprise the set
\begin{equation}
\label{DefZRefin}
 \Z(X,z) \equiv \big\{\, \wt{z} \in \Z(X) \,\big|\, \wt{z} \prec z\,
 \big\} \quad.
\end{equation}

\section{Subtrees} \label{Subtrees}

Let $\B(X)$ be a given tree over $X$. Let $b \in \B(X)$. The set
\begin{equation}
\label{DefSubtree}
 \B(b,X) \equiv \big\{\, b' \in \B(X)\, \big| \, b' \subset b\, \big\}
\end{equation}
will be called the {\it subtree of $b$ with respect to $\B(X)$}. By
definition, $\B(b,X)$ is a tree structure over $b$, and hence an
element of $\M(b)$.

If $b$ is non-refinable, and hence a terminal element in $\B(X)$, then
$\B(b,X) = \{b\}$ is trivial.

\section{Sum, union, extension, reduction, and completion of trees}
\label{OperationsOnTrees}

Let $\B(X)$ be a tree structure over $X$. Consider the elements $X_1,
\ldots, X_{m_1}$ of level 1 in the partition $z(X)$ of $X$, as given
in eq.~(\ref{motiv4}) in section~\ref{Motivation}. For every $X_i$ we
can think of the subtree $\B(X,X_i)$ over $X_i$ with respect to
$\B(X)$. The relation of the subtrees $\B(X,X_i)$, $i=1, \ldots, m_1$,
to the "parent" tree will be described by saying that $\B(X)$ is the
{\it sum} of the trees $\B(X,X_i)$. Now we see how to extend this
definition to tree structures over sets which are not a priori subsets
of a given set: Let $m \in \NN$, let $X_1, \ldots, X_m \neq \emptyset$
be non-empty pairwise disjoint sets, i.e., $X_i \cap X_j = \emptyset$
for $i \neq j$. Let $\B(X_1), \ldots, \B(X_m)$ be tree structures over
$X_1, \ldots, X_m$. Then the set $\sum_{i=1}^m \B(X_i)$, defined by
\begin{equation}
\label{DefSumTree}
 \sum_{i=1}^m \B(X_i) \equiv \left[ \bigcup_{j=1}^m \B(X_j) \right]
 \cup \left\{ \bigcup_{j=1}^m X_j \right\}
\end{equation}
will be called the {\it sum of $\B(X_1), \ldots, \B(X_m)$}. By
construction this is a tree structure over the set
$\bigcup\limits_{j=1}^m X_j$ with subtrees $\B(X_1), \ldots, \B(X_m)$.

Another construction is the {\it union of trees}. This is defined as
follows: Let $\B(X)$ be a tree structure, and let $b \in \B(X)$ be a
terminal but non-primitive element. Then $\# b > 1$. Although $b$ is
not further partitioned in the tree $\B(X)$, we can nevertheless
consider tree structures over $b$ without reference to $\B(X)$. Let
$\B(b)$ be such a tree over $b$. Then we can attach $\B(b)$ to $\B(X)$
by identifying $b \in \B(b)$ with $b \in \B(X)$; the resulting set is
the union $\B(b) \cup \B(X)$, and is again a tree structure which will
be called the {\it union of the trees $\B(b)$ and
$\B(X)$}.

A somewhat related, but more general, concept is the {\it extension of
trees}: Let $\B$ and $\B'$ be two tree structures over the same set
$X$. We will say that $\B'$ is an {\it extension of $\B$} if $\B'
\supsetneqq \B$. A special case of extension is the {\it completion
$\B_c$ of a tree $\B$}: This is defined to be a tree structure $\B_c$
over the same set $X$ as $\B$ that extends $\B$ and is complete, i.e.,
$\B_c$ contains all primitive nodes $\{x\}$ as $x$ runs through
$X$. If $X$ is finite, every tree structure over $X$ admits such a
completion; but clearly, there are many completions $\B_c$ for a given
tree $\B$, which differ in the paths $q(\{x\})$ of the terminal nodes,
see section~\ref{Paths} below.

Yet another construction is the {\it reduction $\B'(X)$ of a tree
$\B(X)$ by a subtree $\B(b,X)$}; this is just the operation inverse to
the union of trees, as defined above: If $b$ is a given node in a
given tree $\B(X)$, we can remove the subtree $\B(b,X)$ from $\B(X)$
by setting
\begin{equation}
\label{DefReducedTree}
 \B'(X) = \big[\, \B(X)\, -\, \B(b)\, \big] \cup \{b\} \quad.
\end{equation}
The set $\B'(X)$ is a tree by construction and is obtained from
$\B(X)$ by simply cutting off the branch containing all further
partitions of $b$, but reattaching $b$ as a terminal element. This
contains an important

\paragraph{Splitting principle} \label{SplittingPrinciple}

Every tree $\B(X)$ can be expressed as the union of any of its
subtrees $\B(b,X)$ with a cutoff tree $\B'(X)$,
\begin{equation}
\label{DefSplitting1}
\B(X) = \B(b,X) \cup \B'(X) \quad,
\end{equation}
where both trees on the right-hand side are subsets of the original
tree and $\B'(X)$ is defined in eq.~(\ref{DefReducedTree}).

\section{Paths in a tree structure} \label{Paths}

We now show that tree structures have a natural partial ordering. To
this end we observe that there exist distinct subsets in a tree
structure which can be totally ordered: Let $\B(X)$ be a given tree
over $X$. Let $b \in \B(X)$. Then we call the set
\begin{equation}
\label{DefPath}
 q(b) \equiv \big\{\, \wt{b} \in \B(X)\; \big|\, \wt{b} \supset b\,
 \big\}
\end{equation}
the {\it path of $b$ in $\B(X)$}. $q(b)$ is certainly non-empty, since
it always contains $X$ and $b$ itself. From the definition of $q(b)$
we see that a total ordering $"\prec"$ on $q(b)$ for all pairs of
elements $(b', b'')$ of $q(b)$ can be defined by setting $b' \prec
b''$ if and only if $b' \supsetneqq b''$. This makes the path $q(b)$ a
totally ordered set, for all nodes $b \in \B(X)$. As a consequence,
the tree $\B(X)$ is a partially ordered set. If $q(b)$ is finite, its
cardinality is a natural number which we denote by $o(b)$,
\begin{equation}
\label{DefPathLength}
 o(b) = \# q(b) \quad,
\end{equation}
and which we shall call the {\it length of the path $q(b)$ in the tree
$\B(X)$}.

Given the natural ordering of the path $q(b)$ as defined above, we
obviously have $b' \prec b$ for all $b' \in q(b)$, by construction of
$q(b)$. Provided that $b \neq X$, it follows that there always exists
an element $b^- \in q(b)$ such that $b^- \prec b$ but $b' \prec b^-$
for all $b' \neq b^-, b$; this distinct element will be called the
{\it predecessor of $b$ in the tree $\B(X)$}. Thus all nodes $b \in
\B(X)$ except for $X$ have a unique predecessor in $\B(X)$; $X$ itself
has no predecessor in $\B(X)$. The predecessor is equal to the
"smallest" node in $\B(X)$ which contains $b$ as a proper subset;
obviously, its own path $q(b^-)$ coincides with the path $q(b)$ just
up to $b$ itself,
\begin{equation}
\label{PathPredecessor}
 q(b) = q(b^-) \cup \{b\} \quad, \quad b \not\in q(b^-) \quad.
\end{equation}

\paragraph{Notation conventions:}

We introduce some notation conventions that will prove convenient in
the sequel.

If $b \in \B(X)$ and $q(b)$ is the path of $b$ in $\B(X)$, then we
denote
\begin{equation}
\label{NotConvent1}
 \dot{q}(b) \equiv q(b)\, -\, \{b\} \quad.
\end{equation}

If $\B(b',X)$ is a subtree of $\B(X)$ and if $b \in \B(b',X)$, then
the path of $b$ in $\B(b',X)$ will be denoted by
\begin{equation}
\label{NotConvent2}
 q_{\B(b',X)}(b) \equiv \big\{\, a \in \B(b',X)\, \big|\, a \supset
 b\, \big\} \quad.
\end{equation}

\section{Partitions compatible with a given tree}
\label{PartitionsCompatible}

Consider a given node $b$ in the tree $\B(X)$. The subtree $\B(b,X)$
defines a distinct set of partitions of $b$ in the following way: Each
distinct partition $z$ is a collection $z= \{b'_1, b'_2, \ldots \}$ of
mutually disjoint subsets $b'_i \subset b$ whose union is $b$ such
that each $b'_i$ is also an element of $\B(b,X)$. Thus, $z \subset
\B(b,X)$. Such a preferred partition of $b$ will be called {\it
compatible with the tree $\B(X)$}. The set of all partitions of $b$
compatible with the tree $\B(X)$ will be denoted by $\zeta(b)$,
\begin{equation}
\label{DefPartCompatible1}
 \zeta(b) \equiv \big\{\, z\in \Z(b)\, \big|\, z \subset \B(b,X)\,
 \big\} \quad.
\end{equation}
If it is necessary to point out that the compatibility is referred to
the given tree $\B(X)$ we shall also use the extended notation
$\zeta(b, \B)$. Similarly, we define $\zeta^*(b)$ to be the set of
nontrivial partitions in $\zeta(b)$, i.e. $\zeta^*(b) = \zeta(b)\,
-\, \{\{b\}\}$.

\subsection{The maximal compatible partition $z_{\tmax}(b)$}
\label{MaxPartition}

The set $\zeta(X)$ contains several distinct partitions: Firstly, the
trivial partition $\{X\}$; and secondly, the partition of $X$ which is
constituted by the set of all terminal nodes in $\B(X)$. Similarly, if
$b \in \B(X)$ is arbitrary, then $\zeta(b)$ contains the trivial
partition $\{b\}$ as well as the partition of $b$ which is constituted
by all terminal elements in the subtree $\B(b,X)$. The latter will be
denoted by $z_{\tmax}(b)$, and will be called the {\it maximal
partition of $b$ in the tree $\B(X)$}. Its elements are those terminal
nodes $b_{\tfin}$ of $\B(X)$ which are also subsets of
$b$. Equivalently we can say that the elements of the maximal
partition $z_{\tmax}(b)$ of $b$ are those terminal elements
$b_{\tfin}$ of $\B(X)$ whose paths $q(b_{\tfin})$ contain $b$,
\begin{equation}
\label{DefMaxPartition}
 z_{\tmax}(b) = \big\{\, b_{\tfin} \in \B(X)\, \big|\, b \in
 q(b_{\tfin}) \big\} \quad.
\end{equation}
Since the maximal partition is defined in terms of terminal elements
it exhibits maximality in the following sense: $z_{\tmax}(b)$ refines
any other partition $z$ of $b$ which is compatible with $\B(X)$,
\begin{equation}
\label{MaxPartition1}
 z \preceq z_{\tmax}(b) \quad \text{for all $z \in \zeta(b,\B)$}
 \quad.
\end{equation}
As a consequence, the number of elements in $z_{\tmax}(b)$ is greater
than or equal to the number of elements in any other $z \in
\zeta(b,\B)$,
\begin{equation}
\label{MaxPartition2}
 \# z \le \# z_{\tmax}(b) \quad \text{for all $z \in \zeta(b,\B)$}
 \quad.
\end{equation}

If $b$ is terminal, then the maximal partition is the trivial
partition, $z_{\tmax}(b) = \{b\}$, as follows from
eq.~(\ref{DefMaxPartition}). In this case $\# z_{\tmax}(b) = 1$.

If $b$ is not a terminal node then $z_{\tmax}(b) \in \zeta^*(b)$. If
$b = X$, then the union of all paths $q(b_{\tfin})$ with $b_{\tfin}
\in z_{\tmax}(X)$ renders the whole tree structure $\B(X)$,
\begin{equation}
\label{MaxPartition3}
 \bigcup_{b_{\tfin} \in z_{\tmax}(X)} q(b_{\tfin}) = \B(X) \quad.
\end{equation}

The maximal partition is related to the concept of reduced trees,
section~\ref{OperationsOnTrees}, and the concept of the set of
partitions $\zeta(b)$ compatible with $\B(X)$, in the following way:
\begin{theorem} \label{CompatiblePartitionsAndReducedTrees}
Let $\B(X)$ be a given tree over $X$. The set $\zeta(X)$ of partitions
of $X$ compatible with $\B(X)$ is comprised of the maximal partitions
$z'_{\tmax}(X)$ of $X$ with respect to $\B'(X)$, where $\B'(X)$ ranges
through all reduced trees (\ref{DefReducedTree}) associated with
$\B(X)$.
\end{theorem}
\begin{proof}
Let $z \in \zeta(X)$, then all elements $b_1, b_2, \ldots$ of $z$ are
elements of $\B(X)$. Consider the union of paths
\begin{equation}
\label{UnionOfPaths1}
 \bigcup_i q(b_i) \equiv \B'(X) \quad,
\end{equation}
where $q(b_i)$ are the paths of $b_i$ in $\B(X)$. By construction, the
right-hand side $\B'(X)$ is a tree structure, and is also a reduction
of the original tree $\B(X)$, with maximal partition $z_{\tmax}(X,\B')
=z$. Conversely, let $\B'(X)$ be a reduction of $\B(X)$ with maximal
partition $z'_{\tmax}(X)$; then all elements $b' \in z'_{\max}(X)$ lie
in $\B(X)$ by definition of a reduced tree; hence $z'_{\tmax}(X)$ is
compatible with $\B(X)$.
\end{proof}

\subsection{The minimal compatible partition $z_{\tmin}(b)$}
\label{MinPartition}

Let $b$ be a given non-terminal node in the tree. The partition of $b$
that is obtained by stepping to the next level in the tree will be
called the {\it minimal partition $z_{\tmin}(b)$ of $b$ in
$\B(X)$}. The elements $b'$ in the minimal partition are uniquely
characterised by the feature that they all have the node $b$ as their
predecessor, and there are no further nodes $b'$ which have this
predecessor. We can therefore write
\begin{equation}
\label{DefMinPartition}
 z_{\tmin}(b) = \big\{\, b' \in \B(X)\, \big| \, \left(b'\right)^- = b
 \, \big\} \quad.
\end{equation}
$z_{\tmin}(b)$ has another minimal property which can be alternatively
used to define it as a set: The minimal partition $z_{\tmin}(b)$ of
$b$ is uniquely characterised by the fact that it is refined by any
non-trivial partition of $b$ compatible with $\B(X)$
\begin{equation}
\label{MinPartition1}
 z_{\tmin}(b) \preceq z \quad \text{for all $z \in \zeta^*(b,\B)$}
 \quad,
\end{equation}
and as a consequence contains the least number of elements,
\begin{equation}
\label{MinPartition2}
 \# z_{\tmin}(b) \le \# z \quad \text{for all $z \in \zeta^*(b,\B)$}
 \quad.
\end{equation}
This follows immediately from the definition~(\ref{DefMinPartition})
of $z_{\tmin}$.

The above definition~(\ref{DefMinPartition}) or, alternatively,
eq.~(\ref{MinPartition1}), is meaningful only when $b$ is not a
terminal element of the given tree $\B(X)$. If $b$ is terminal we
define the minimal partition to be the trivial partition,
$z_{\tmin}(b) = \{b\}$. In this case $\# z_{\tmin}(b)~=~1$.

Let $b$ be any non-terminal element of $\B(X)$. Given the minimal
partition $z_{\tmin}(b)$ of $b$ in $\B(X)$, we can split the subtree
$\B(b,X)$ accordingly into a sum of subtrees,
\begin{equation}
\label{MinPartition3}
 \B(b,X) = \sum_{b' \in z_{\tmin}(b)} \B(b',b) \quad.
\end{equation}
We shall make use of this fact frequently.

The number of elements in the minimal partition $z_{\tmin}(b,\B)$ of
$b$ in the given tree $\B(X)$ will be denoted as
\begin{subequations}
\begin{equation}
\label{NodeCharacter1}
 m(b) \equiv \# z_{\tmin}(b,\B) \quad.
\end{equation}
Similarly, the number of elements of $b$ regarded as a set will be
denoted by $n(b)$,
\begin{equation}
\label{NodeCharacter2}
 n(b) \equiv \# b \quad.
\end{equation}
\end{subequations}
These quantities pertain to the nodes $b \in \B(X)$ in a specific way
and will play a crucial role in what follows. We must have
\begin{equation}
\label{NodeCharacter3}
 n(b) = \sum_{a \in z_{\tmin}(b)} n(a) \quad.
\end{equation}
For every $b \in \B(X)$ the following inequality holds:
\begin{equation}
\label{NodeCharacter4}
 1 \le m(b) \le n(b) \quad.
\end{equation}
Furthermore, if $b \neq X$, then $b$ has a unique predecessor $b^-$,
whose minimal partition $z_{\tmin}(b^-)$ has $m(b^-)$ elements, one of
which is just $b$. Each of the nodes in $z_{\tmin}(b^-)$ contains at
least one element, and there are $m(b^-)-1$ nodes apart from $b$;
hence
\begin{equation}
\label{NodeCharacter5}
 n(b^-) \ge m(b^-) - 1 + n(b) \quad,
\end{equation}
or
\begin{equation}
\label{NodeCharacter6}
 m(b^-) - 1 \le n(b^-) - n(b) \quad.
\end{equation}

\subsection{Trees reduced by a partition}

By means of the concept of a partition compatible with a given tree we
can introduce a generalization of the idea of reduced trees as given
in eq.~(\ref{DefReducedTree}) in section~\ref{OperationsOnTrees}:

Let $\B(X)$ be a given tree over the set $X$. Let $z \in \zeta(X,\B)$
be a partition of $X$ compatible with the tree $\B(X)$. Then we can
construct a new tree $\B(z)$ as follows: For each $b \in z$, we remove
the subtree $\B(b,X)$ from $\B(X)$ but reattach $b$ as a terminal
element; this is just the proper generalization of
eq.~(\ref{DefReducedTree}). The set so obtained is again a tree by
construction:
\begin{definition}[Tree reduced by a partition]
\label{DefTreeReducedByPartition}
The tree structure $\B(z)$ defined by
\begin{equation}
\label{EqDefTreeReducedByPartition}
 \B(z) \equiv \left[\, \B(X) - \bigcup\limits_{b \in z} \B(b,X)\,
 \right] \cup \Bigg[\, \bigcup\limits_{b \in z} \{b\} \, \Bigg] \quad,
\end{equation}
is called the {\em tree $\B(X)$ reduced by the partition $z \in
\zeta(X,\B)$}.
\end{definition}
If $\wt{b}$ is an element of the reduced tree $\B(z)$, then the
subtree of $\wt{b}$ in the reduced tree will be denoted by
$\B(\wt{b},z)$.

\section{Amount functions} \label{Amountunctions}

From now on we explicitly assume that $X$ is a finite set. As a
consequence, the quantities $m(b)$ and $n(b)$ are always finite
natural numbers.

Let $b \in \B(X)$ with $b \neq X$. Then $b^-$ exists, and the number
of elements in $z_{\tmin}(b^-)$ is $m(b^-)$. Now we think of $b$ as
being distinct in the set of elements $b'$ comprising
$z_{\tmin}(b^-)$. Suppose we are presented the set $z_{\tmin}(b^-) =
\{b'_1, \ldots, b'_{m(b^-)} \}$, as in the scenario laid out in
section~\ref{Motivation}, and we are asked to find out which of the
$b'_i$ is the distinct one. Presuming that no optimized search
strategy is employed we have to expend at most $(m(b^-)-1)$ questions
in order to fulfill our task.

We can now extend this reasoning to the whole path $q(b)$: $b^-$ is
distinct in the set of all $b''$ comprising the minimal decomposition
$z_{\tmin}(b^{2-})$, where $b^{2-}$ denotes the predecessor of $b^-$
in $\B(X)$. In order to determine $b^-$ amongst the $m(b^{2-})$
elements of $z_{\tmin}(b^{2-})$ we have to expend at most
$(m(b^{2-})-1)$ questions. We can continue in this way up the whole
path $q(b)$ until no predecessor $b^{(k+1)-}$ exists any longer, in
other words, $b^{k-} = X$. The maximum number of questions to
determine the distinct node $b \in \B(X)$ we were seeking out is
therefore the sum of all these contributions,
\begin{equation}
\label{DefAmountOfB}
 e(b) \equiv \sum\limits_{a\in \dot{q}(b)\}} \big[\, m(a) -1\, \big] =
 \left[\, \sum\limits_{a\in \dot{q}(b)} m(a)\, \right] - o(b) + 1
 \quad,
\end{equation}
where the length of the path $o(b)$ was defined in
eq.~(\ref{DefPathLength}).
\begin{definition}[Amount of a node] \label{AmountOfNode}
 The quantity $e(b)$ in eq.~(\ref{DefAmountOfB}) will be called {\em
the amount of $b$} in the tree $\B(X)$.
\end{definition}
When emphasizing the fact that the amount is dependent on the
underlying tree we shall also use the notation $e_{\B}(b)$.

\paragraph{Remark:}
In eq.~(\ref{DefAmountOfB}), the element $b$ is excluded from
summation, since $a \in \dot{q}(b)$ only. If $b$ is terminal in
$\B(X)$ then $m(b)=1$; in this case we can trivially extend the sums
in (\ref{DefAmountOfB}) to range over the whole path $q(b)$, since the
additional contribution $m(b)-1$ is zero on account of $m(b)=1$. It is
then possible to write the path amount (\ref{DefAmountOfB}) as
\begin{equation}
\label{DefAmountOfBRemark1}
 e(b) = \sum\limits_{a\in q(b)} \bigg[ m(a) -1 \bigg] \quad \text{for
 $b=b_{\text{fin}} \in z_{\tmax}(X)$} \quad.
\end{equation}
We shall frequently make use of this convention.

Now let $z \in \zeta(X)$ be an arbitrary partition of $X$ compatible
with the tree $\B(X)$. Then every element $b \in z$ has the uniquely
defined path $q(b) \subset \B(X)$. Hence it makes sense to sum up the
amounts $e_{\B}(b)$ of each $b$:
\begin{definition}[Total amount of partition] \label{TotalAmountOfPartition}
The quantity $G(z)$, defined by
\begin{equation}
\label{DefAmountOfPartition}
 G(z) \equiv \sum\limits_{b \in z} e_{\B(X)}(b) = \sum\limits_{b \in
 z}\; \sum\limits_{a \in \dot{q}(b)} \big[\, m(a) -1\,\big] \quad,
\end{equation}
is called the {\em total amount of $z \in \zeta(X)$} with respect to
the tree $\B(X)$. If $z$ contains only one element we define the
associated amount to be $G(z) = 0$.
\end{definition}
When emphasizing the fact that the total amount is dependent on the
underlying tree structure we shall also denote $G(z) \equiv
G_{\B}(z)$.

\vspace{1ex}

Now consider the maximal partition $z_{\tmax}(X)$ of $X$ in $\B(X)$:
From
\begin{equation}
\label{TotalAmountOfTree}
\begin{aligned}
 G_{\B(X)} & \equiv G(z_{\tmax}(X)) = \sum\limits_{b \in z_{\tmax}(X)}
 e_{\B(X)}(b) = \\
 & = \sum\limits_{a \in \B(X) - z_{\tmax}(X)} \big[\, m(a) -1\, \big]
 \quad
\end{aligned}
\end{equation}
we see that in this case we sum over all but the terminable elements
$b \in \B(X)$; hence the total amount for the maximal partition of $X$
in $\B(X)$ is dependent on $\B(X)$ only; this is reflected in our
notation. The sum $G_{\B(X)}$ therefore defines a map from the set of
all tree structures over $X$ into the natural numbers,
\begin{equation}
\label{DefAmountFunction}
 G:\; \left\{ \begin{array}{rcl} \M(X) & \rightarrow & \NN \\ \B(X) &
 \mapsto & G_{\B(X)} \end{array} \quad. \right.
\end{equation}
\begin{definition}[Total amount of tree. Amount function]
The quantity $G_{\B(X)}$ is called the {\em total amount of the tree
structure $\B(X)$}. The map $G$, as defined in
eq.~(\ref{DefAmountFunction}), is called the {\em amount function} on
$\M(X)$.
\end{definition}
\paragraph{Remark 1:}
If the tree $\B(X) = \{X\}$ is trivial we again define the associated
total amount to be zero, $G_{\B(X)} = 0$.
\paragraph{Remark 2:}
The total amount $G(z)$ with respect to the partition $z \in \zeta(X)$
compatible with the given tree $\B(X)$, defined in
eq.~(\ref{DefAmountOfPartition}), is equal to the total amount
$G_{\B(z)}$ of the reduced tree $\B(z)$, which is obtained by reducing
$\B(X)$ via $z$ according to
definition~\ref{DefTreeReducedByPartition}. We can use this fact to
emphasize that the total amount of the partition $G(z)$ is dependent
on the underlying tree structure $\B(X)$,
\begin{equation}
\label{AmountOfReducedTree}
 G(z) \equiv G_{\B(z)} \quad.
\end{equation}

\begin{proposition}[Inequalities]
For every $b \in \B(X)$ the following inequalities hold:
\begin{equation}
\label{Inequalities1}
 o(b) - 1 \le e(b) \le n(X) - n(b)  \le n(X) - 1 \quad.
\end{equation}
\end{proposition}
\begin{proof}
Set $o(b) = \kappa$ and $q(b) = \{ \beta_1, \ldots, \beta_{\kappa}\}$,
with $\beta_1 = X$ and $\beta_{\kappa} = b$. Then $\dot{q}(b) = \{
\beta_1, \ldots, \beta_{\kappa-1} \}$, and we must have
\begin{equation}
\label{proofIE1}
 e(b) = \sum\limits_{j=1}^{\kappa-1} \big[\, m(\beta_j) - 1\, \big] =
 \sum\limits_{j=2}^{\kappa} \big[\, m(\beta_{j-1}) - 1 \, \big] \quad.
\end{equation}
For all $j \in \{1, \ldots, \kappa-1 \}$ we must have $m(\beta_j) \ge
2$. If this is inserted into eq.~(\ref{proofIE1}) we obtain the first
inequality in (\ref{Inequalities1}). If eq.~(\ref{NodeCharacter6}) is
inserted into (\ref{proofIE1}) we find
\begin{equation}
\label{proofIE2}
\begin{aligned}
 e(b) & \le \sum\limits_{j=2}^{\kappa} \big[\, n(\beta_{j-1}) -
 n(\beta_j)\, \big] = \sum\limits_{j=1}^{\kappa-1} n(\beta_j) -
 \sum_{j=2}^{\kappa} n(\beta_j) = \\
 & = n(\beta_1) - n(\beta_{\kappa}) = n(X) - n(b) \quad.
\end{aligned}
\end{equation}
This yields the second inequality in (\ref{Inequalities1}). The last
inequality follows trivially from the fact that $n(b) \ge 1$.
\end{proof}

\section{Induced Partitions}  \label{InducedPartitions}

Let $\B(X)$ be a given tree over $X$. Let $b \in \B(X)$. For every $z
\in \zeta(X)$ we can introduce the intersection
\begin{equation}
\label{DefInducedPartition}
 \sigma(z,b) \equiv z \cap \B(b,X) \quad.
\end{equation}
$\sigma(z,b)$ can be empty if all elements in $z$ are "coarser" than
$b$, i.e., $b \subsetneqq b'$ for precisely one $b' \in z$, and has
zero intersection with the rest. If $\sigma(z,b)$ is non-empty, it is
a partition of $b$ compatible with the subtree $\B(b,X)$, as we show
now:
\begin{theorem}[Induced partitions] \label{TheoInducedPartitions}
\rule{0pt}{0ex}
\begin{description}
\item[(A)] If $\sigma(z,b) \neq \emptyset$ then $\sigma(z,b) \in
\zeta(b,\B)$.
\item[(B)] Conversely, if $\wt{z} \in \zeta(b,\B)$, then there exists
$z \in \zeta(X)$ such that $\sigma(z,b) = \wt{z}$.
\item[(C) Definition: ] If $\sigma(z,b)$ is non-empty it is called the
{\it partition of $b$ induced by $z$}.
\end{description}
\end{theorem}
\begin{proof}
Assume that $\sigma(z,b) \neq \emptyset$, then $\sigma(z,b) = \{b_1,
b_2, \ldots\}$, where $b_i \in \B(b,X)$. Now let $\b{b} \in z -
\sigma(z,b)$, then $\b{b}$ cannot intersect $b$: For, $\b{b}$ lies in
$\B(X)$ but not in the subtree $\B(b,X)$ by assumption; hence if it
intersects $b$ then it must contain $b$ properly, $\b{b} \supsetneqq
b$. But then it also contains all $b_i$ as subsets, which contradicts
the fact that the $b_i, \b{b}$ are mutually disjoint. Thus, $\b{b}
\cap b = \emptyset$. It follows that the union of all $\b{b} \in z -
\sigma(z,b)$ has no intersection with $b$. As a consequence, the union
of all $b_i$ must be equal to $b$, since $z$ is a partition of
$X$. Furthermore, the $b_i$ are mutually disjoint, and lie in
$\B(b,X)$, from which it follows that $\{b_1, b_2, \ldots\} \in
\zeta(b,X)$. This proves (A).

Let $\wt{z} \in \zeta(b,\B)$ be given. Now represent $\B(X)$ as a
union $\B(X) = \B'(X) \cup \B(b,X)$ of trees as in
section~\ref{OperationsOnTrees}, where the reduced tree $\B'(X)$ is
given in eq.~(\ref{DefReducedTree}). Then the maximal partition
$z_{\tmax}(X,\B')$ of $X$ in the reduced tree contains $b$ as an
element. We now define a new partition by removing $b$ from
$z_{\tmax}(X,\B')$ and replacing it by the set of elements in
$\wt{z}$,
\begin{equation}
\label{DefNewPartition}
 z \equiv \Big[\, z_{\tmax}(X,B')\, - \{b\}\, \Big] \cup
 \wt{z} \quad.
\end{equation}
The set $z$ so defined is obviously a partition of $X$ compatible with
$\B(X)$, hence $z \in \zeta(X,B)$, and, by construction, $z \cap
\B(b,X) = \wt{z}$. This proves (B).
\end{proof}

The concept of induced partitions is linked to the idea of refinements
of partitions:
\begin{proposition}[Refinement of partitions] \label{RefinementOfPartitions}
Let $\B(X)$ be a tree over $X$. Let $z, z' \in \zeta(X)$ with $z \prec
z'$. Then
\begin{description}
\item[(A) ] $z - (z\cap z') \neq \emptyset$.
\item[(B) ] $\sigma(z',b) \in \zeta^*(b)$ for all $b \in z -
(z \cap z')$, whereas $\sigma(z',b) = \{b\}$ is the trivial partition
for all $b \in z \cap z'$.
\item[(C) ]
\begin{equation}
\label{EqRefinementPartitions}
 z' - (z' \cap z) = \bigcup\limits_{b \in z - (z
 \cap z')} \sigma(z', b) \quad.
\end{equation}
\end{description}
\end{proposition}
\begin{proof}
Since $z'$ is a refinement of $z$, any element $b'$ of $z'$ is
contained in some element $b$ of $z$ as a subset, $b' \subset b$. $z
\cap z'$ contains all elements which are not partitioned under the
refinement $z \rightarrow z'$. This means that for all $b \in z \cap
z'$, $z' \cap \B(b,X) = \{b\}$, hence $\sigma(z',b)$ is the trivial
partition of $b$. This proves the second statement in (B). On the
other hand, if $b \in z - (z \cap z')$, then $b$ is undergoing a
proper partition under the refinement $z \rightarrow z'$. This implies
that $\sigma(z',b) \in \zeta^*(b)$, thus proving the first statement
in (B). Since $z'$ is refined there must exist at least one element of
$z$ that undergoes a proper partition, which says that $z - (z \cap
z')$ cannot be empty, hence (A). Finally,
\begin{equation}
\label{EqRefPartProof1}
 z'-(z' \cap z) = z' \cap \Big[\, \bigcup\limits_{b \in z-(z \cap z')}
 \B(b,X) \Big] = \bigcup\limits_{b \in z-(z \cap z')} z' \cap \B(b,X)
 \quad,
\end{equation}
but $z' \cap \B(b,x) = \sigma(z',b)$, hence
eq.~(\ref{EqRefinementPartitions}) follows.
\end{proof}


The splitting theorem describes the behaviour of total amount functions
of reduced trees $\B(z)$ and $\B(z')$, where $z'$ is a refinement of
$z$:
\begin{theorem}[Splitting theorem] \label{SplittingLemma}
Let $\B(X)$ be a given tree over $X$. Let $z,z' \in \zeta(X)$ with $z
\prec z'$. Let $\B(z)$ and $\B(z')$ be the corresponding reduced
trees. Then
\begin{equation}
\label{EqSplitLemma1}
 G(z') - G(z) = \sum\limits_{b \in z-(z\cap z')} \Big[\, \#
 \sigma(z',b) -1\, \Big] \cdot e_{\B(z)}(b) + \sum\limits_{b \in
 z-(z\cap z')} G_{\B(b,z')} \quad.
\end{equation}
\end{theorem}
\begin{proof}
Using eq.~(\ref{AmountOfReducedTree}) we have
\begin{equation}
\label{SplLemmaProof1}
\begin{aligned}
 G(z') & = \sum\limits_{b' \in z'} e_{\B(z')}(b') = \sum\limits_{b'
 \in z' \cap z} e_{\B(z')}(b') + \sum\limits_{b' \in z'-(z' \cap z)}
 e_{\B(z')}(b') = \\
 & = \sum\limits_{b' \in z' \cap z} e_{\B(z')}(b') + \sum\limits_{b'
 \in z'-(z' \cap z)}\; \sum\limits_{a' \in \dot{q}_{\B(z')}(b')}
 \big[\, m(a') -1\, \big] = \\
 & = \sum\limits_{b' \in z'\cap z} e_{\B(z')}(b') + {\rm ZS} \quad,
\end{aligned}
\end{equation}
where
\begin{equation}
\label{SplLemmaProof2}
 {\rm ZS} = \sum\limits_{b' \in z'- (z \cap z')}\; \sum\limits_{a' \in
 \dot{q}_{\B(z')}(b')} \big[\, m(a') -1\, \big] \quad.
\end{equation}
With the help of eq.~(\ref{EqRefinementPartitions}) we can split the
sums in ${\rm ZS}$ further:
\begin{equation}
\label{SplLemmaProof3}
 {\rm ZS} = \sum\limits_{b \in z- (z\cap z')} \sum\limits_{b' \in
 \sigma(z',b)} \sum\limits_{a' \in \dot{q}_{\B(z')}(b')} \big[\, m(a')
 - 1\, \big] \quad.
\end{equation}
Since $b \in z-(z\cap z')$ and $b' \in \sigma(z',b)$, we have $b \in
\dot{q}_{\B(z')}(b')$. But
\begin{subequations}
\label{SplLemmaProof4}
\begin{equation}
 \big\{\, a' \in \dot{q}_{\B(z')}(b')\, \big|\, a' \subset b\, \big\}
 = \dot{q}_{\B(b,z')}(b') \quad, \label{SpLePr4a}
\end{equation}
and
\begin{equation}
 \big\{\, a' \in \dot{q}_{\B(z')}(b')\, \big|\, a' \supsetneqq b\,
 \big\} = \dot{q}_{\B(z)}(b) \quad, \label{SpLePr4b}
\end{equation}
\end{subequations}
for all $b' \in \sigma(z',b)$ and $b \in z- (z\cap z')$. Hence we can
write
\begin{equation}
\label{SplLemmaProof5}
\begin{aligned}
 \dot{q}_{\B(z')}(b') & = \big\{\, a' \in \dot{q}_{\B(z')}(b')\,
 \big|\, a' \supsetneqq b\, \big\} \cup \big\{\, a' \in
 \dot{q}_{\B(z')}(b')\, \big|\, a' \subset b\, \big\} = \\
 & = \dot{q}_{\B(z)}(b)\, \cup\, \dot{q}_{\B(b,z')}(b') \quad.
\end{aligned}
\end{equation}
This yields
\begin{equation}
\label{SplLemmaProof6}
\begin{aligned}
 {\rm ZS} & = \sum\limits_{b \in z-(z\cap z')}\, \sum\limits_{b' \in
 \sigma(z',b)} \bigg\{\; \sum\limits_{a' \in \dot{q}_{\B(z)}(b)}
 \big[\, m(a')-1\, \big]\; + \\
 & + \sum\limits_{a' \in \dot{q}_{\B(z',b)}(b')} \big[\, m(a') -1\,
 \big]\; \bigg\} = \\
 & = \sum\limits_{b \in z-(z\cap z')}\, \sum\limits_{b' \in
 \sigma(z',b)} e_{\B(z)}(b) + \sum\limits_{b \in z-(z\cap z')}\,
 \sum\limits_{b' \in \sigma(z',b)} e_{\B(z',b)}(b') = \\
 & = \sum\limits_{b \in z-(z\cap z')} \# \sigma(z',b) \cdot
 e_{\B(z)}(b) + \sum\limits_{b \in z-(z\cap z')}\, \sum\limits_{b' \in
 \sigma(z',b)} e_{\B(z',b)}(b') \quad,
\end{aligned}
\end{equation}
where we have used eq.~(\ref{DefAmountOfB}) for $e_{\B(z)}(b)$. We now
insert ${\rm ZS}$ into eq.~(\ref{SplLemmaProof1}) for $G(z')$:
\begin{equation}
\label{SplLemmaProof7}
\begin{aligned}
 G(z') & = \sum\limits_{b' \in z'\cap z} e_{\B(z')}(b') +
 \sum\limits_{b \in z-(z \cap z')} \# \sigma(z', b) \cdot
 e_{\B(z)}(b)\; + \\
 & + \sum\limits_{b \in z-(z\cap z')}\, \sum\limits_{b' \in
 \sigma(z',b)} e_{\B(z',b)}(b') = \\
 & = \sum\limits_{b \in z\cap z'} e_{\B(z)}(b) + \sum\limits_{b \in
 z-(z\cap z')} e_{\B(z)}(b) + \\
 & + \sum\limits_{b \in z-(z\cap z')} \big[\, \# \sigma(z',b) -1\,
 \big] \cdot e_{\B(z)}(b) + \sum\limits_{b \in z-(z\cap z')}
 G_{\B(z',b)} \quad
\end{aligned}
\end{equation}
where we have used the fact that, for elements $b \in z\cap z'$, the
amount $e_{\B(z')}(b)$ of the path of $b$ in the tree $\B(z')$ is
equal to the amount $e_{\B(z)}(b)$ of the path of $b$ in the tree
$\B(z)$. Then the first two terms on the right-hand side of the last
equation combine to give
\begin{equation}
\label{SplLemmaProof8}
 \sum\limits_{b \in z\cap z'} e_{\B(z')}(b) + \sum\limits_{b \in
 z-(z\cap z')} e_{\B(z)}(b) = \sum\limits_{b \in z} e_{\B(z)}(b) =
 G(z) \quad.
\end{equation}
If eq.~(\ref{SplLemmaProof8}) is inserted into
eq.~(\ref{SplLemmaProof7}) we obtain eq.~(\ref{EqSplitLemma1}).
\end{proof}

We see from eq.~(\ref{EqSplitLemma1}) that there are two contributions
to the difference in the total amounts: The first one links the
amounts $e_{\B(z)}(b)$ of the paths $q(b)$ of $b$ in $\B(z)$ to the
"degree of splitting"  $\# \sigma(z',b)$ of the set $b$ under the
refinement $z \rightarrow z'$; the second one is the sum of all
amounts $G_{\B(z',b)}$ of the subtrees $\B(b,z')$ of the larger tree
$\B(z')$.
\begin{corollary} \label{SplittingLemmaCorollary}
For the special case $z = z_{\tmin}(X)$ and $z' = z_{\tmax}(X)$ we
have
\begin{equation}
\label{EqSplittingLemmaCorollary1}
 G_{\B(X)} = \big[\, m(X) - 1\, \big] \cdot \sum\limits_{b \in
 z_{\tmin}(X)} \# z_{\tmax}(b) + \sum\limits_{b \in z_{\tmin}(X)}
 G_{\B(b,X)} \quad.
\end{equation}

If the tree $\B(X)$ is complete then
\begin{equation}
\label{EqSplittingLemmaCorollary2}
 G_{\B(X)} = \big[\, m(X) -1\,\big] \cdot n(X) + \sum\limits_{b \in
 z_{\tmin}(X)} G_{\B(b,X)} \quad.
\end{equation}
\end{corollary}
\begin{proof}
For $z = z_{\tmin}(X)$, $z' = z_{\tmax}(X)$ we have $G(z) = m(X) \cdot
[m(X)-1]$, as follows from eq.~(\ref{DefAmountOfPartition}), and
$G(z') = G_{\B(X)}$, since $\B(z_{\tmax}(X)) = \B(X)$ is the total
tree $\B(X)$. Furthermore, for $b \in z- (z \cap z')$ we have
$\sigma(z',b) = z_{\tmax}(b) \in \zeta(b)$, and $G_{\B(b,z')}=
G_{\B(b,X)}$. Then eq.~(\ref{EqSplitLemma1}) gives
\begin{equation}
\label{EqSplLemCoProof1}
\begin{aligned}
 G_{\B(X)} & = m(X) \cdot \big[\, m(X) -1\, \big]\; + \\
 & + \sum\limits_{b \in z-(z\cap z')} \big[\, \# z_{\tmax}(b)-1\,\big]
 \cdot \underset{= m(X)-1}{\underbrace{e_{\B(X)}(b)}}\; +
 \sum\limits_{b\in z-(z\cap z')} G_{\B(b,X)}\; = \\
 & = \big[\, m(X)-1\, \big] \cdot \bigg\{\, m(X) + \sum\limits_{b \in
 z-(z\cap z')} \big[\,\# z_{\tmax}(b)-1\, \big] \bigg\}\; + \\
 & + \sum\limits_{b \in z-(z\cap z')} G_{\B(b,X)} \quad.
\end{aligned}
\end{equation}
However, in all sums we can extend the range of $b$ to take values in
$z\cap z'$ as well; for such an element $b$, $\# z_{\tmax}(b)=1$, and
$G_{\B(b,X)}=0$. Thus, $b$ can be allowed to run over the whole set
$z=z_{\tmin}(X)$,
\begin{equation}
\label{EqSplLemCoProof2}
\begin{aligned}
 G_{\B(X)} & = \big[\, m(X)-1\, \big] \cdot \bigg\{\, m(X) +
 \sum\limits_{b \in z_{\tmin}(X)} \big[\,\# z_{\tmax}(b)-1\, \big]
 \bigg\}\; + \\
 & + \sum\limits_{b \in z_{\tmin}(X)} G_{\B(b,X)} \quad.
\end{aligned}
\end{equation}
The first and the third contribution in curly brackets cancel each
other; thus, we arrive at eq.~(\ref{EqSplittingLemmaCorollary1}).

If the tree is complete then the maximal partition $z_{\tmax}(X)$ is
complete, i.e.,
\begin{equation}
\label{EqSplLemProof3}
 z_{\tmax}(X) = \left\{\, \{x_1\}, \, \{x_2\}, \ldots, \, \right\}
 \quad,
\end{equation}
where $x_i$ are the elements of $X$. In this case, each of the maximal
partitions $z_{\tmax}(b)$ for $b \in z_{\tmin}(X)$ is complete, so that
\begin{equation}
\label{EqSplLemProof4}
 \# z_{\tmax}(b) = \# b \quad \text{for all $b \in z_{\tmin}(X)$}
 \quad.
\end{equation}
Consequently,
\begin{equation}
\label{EqSplLemProof5}
 \sum\limits_{b \in z_{\tmin}(X)} \# z_{\tmax}(b) = \# X = n(X) \quad,
\end{equation}
from which eq.~(\ref{EqSplittingLemmaCorollary2}) follows.
\end{proof}

\section{The tree function $E_{\B(X)}$} \label{TreeFunction}

The next theorem will be the first main statement about the properties
of amount functions, in that it expresses the amount of a tree as a
function of the pairs of numbers $\big( n(b), m(b) \big)$ at every
node $b \in \B(X)$. To formulate this we need to define a new
quantity:
\begin{definition}[Tree function] \label{DefTreeFunction}
Let $\B(X)$ be a tree structure over $X$. The {\em tree function}
$E_{\B(X)}$ of the tree $\B(X)$ is defined to be
\begin{equation}
\label{EqDefTreeFunction}
 E_{\B(X)} \equiv \sum\limits_{b \in \B(X)} n(b) \cdot \big[\,
 m(b)-1\, \big] \quad,
\end{equation}
where the sum runs over all nodes in the tree.
\end{definition}
\begin{theorem}[Tree function and total amount]
\label{TreeFunctionAndTotalAmount}
Let $\B(X)$ be a tree structure over $X$.
\begin{description}
\item[(A)] For a general tree,
\begin{equation}
\label{EqTreeFunctionAndTotalAmount1}
 E_{\B(X)} = \sum\limits_{b \in z_{\tmax}(X)} n(b) \cdot e_{\B(X)}(b)
 \quad.
\end{equation}
\item[(B)] If $\B(X)$ is complete then
\begin{equation}
\label{EqTreeFunctionAndTotalAmount2}
 E_{\B(X)} = G_{\B(X)} \quad.
\end{equation}
\end{description}
\end{theorem}
These results say that, for a complete tree, the tree function
coincides with the total amount in the tree, whereas if the tree is
incomplete, then the tree function renders a weighted sum of the path
amounts $e_{\B(X)}(b)$, the weights being equal to the cardinality
$n(b)$ of the terminal elements $b \in z_{\tmax}(X)$ in the incomplete
tree.

\begin{proof}
We first prove (A) by induction with respect to $n \equiv \# X$: For
$n=1$, both left-hand side (LHS) and right-hand side (RHS) are zero
and hence agree.

For $n=2$, there are only two possible trees: Either, $\B(X)= \{X\}$
is the trivial tree, in which case $z_{\tmax}(X) = \{X\}$,
$e_{\B(X)}(X) =0$, and $m(X)=1$, so that again, LHS and RHS agree to
give zero. Or, $\B(X)= \{X, \{x_1\}, \{x_2\} \}$. In this case, the
LHS is equal to
\begin{equation}
\label{TreeFunctionAndTotalAmountProof1}
 E_{\B(X)} = 2\cdot (2-1) + 1\cdot(1-1) + 1\cdot (1-1) = 2 \quad.
\end{equation}
On the RHS, $z_{\tmax}(X) = \{ \{x_1\}, \{x_2\} \}$, $e_{\B(X)}(b)=1$
and $n(b)=1$ for $b \in z_{\tmax}(X)$, so that the RHS also yields
$2$.

We now perform the induction: We assume that
eq.~(\ref{EqTreeFunctionAndTotalAmount1}) holds for all possible sets
$X$ with $\# X \in \{1, \ldots, n-1\}$. We shall prove that
eq.~(\ref{EqTreeFunctionAndTotalAmount1}) is valid for sets $X$ with
$\# X=n$ as well. If $\B(X)=\{X\}$ is trivial then
eq.~(\ref{EqTreeFunctionAndTotalAmount1}) holds trivially as
before. Thus we can assume that the tree is nontrivial, which implies
that the set $X$ is properly split in the tree, hence $\# z_{\tmin}(X)
> 1$. As a consequence, the cardinality of each $a \in z_{\tmin}(X)$
must be smaller than that of $X$,
\begin{equation}
\label{TreeFunctionAndTotalAmountProof2}
 \# a < \# X = n \quad \text{for all $a \in z_{\tmin}(X)$} \quad.
\end{equation}
Now we decompose $X$ into a sum of subtrees $\B(a,X)$, where $a \in
z_{\tmin}(X)$, as in eq.~(\ref{DefSumTree}). Then the LHS of
eq.~(\ref{EqDefTreeFunction}) can be split into
\begin{equation}
\label{TreeFunctionAndTotalAmountProof3}
 E_{\B(X)} = n(X) \cdot \big[\, m(X)-1\, \big] + \sum\limits_{a \in
 z_{\tmin}(X)}\; \sum\limits_{b \in \B(a,X)} n(b) \cdot \big[\,
 m(b)-1\, \big] \quad.
\end{equation}
For each of the sums $\sum_{b \in \B(a,X)}$ on the RHS of
(\ref{TreeFunctionAndTotalAmountProof3}), the induction assumption
applies,
\begin{equation}
\label{TreeFunctionAndTotalAmountProof4}
 \sum\limits_{b \in \B(a,X)} n(b) \cdot \big[\, m(b)-1\, \big] =
 E_{\B(a,X)} = \sum\limits_{b \in z_{\tmax}(a)} n(b) \cdot
 e_{\B(a,X)}(b) \quad,
\end{equation}
where the maximal partition $z_{\tmax}(a)$ of $a$ refers to the
subtree $\B(a,X)$, but is clearly the same as with respect to the full
tree $\B(X)$. Now assume that $b \in z_{\tmax}(a)$ for some $a \in
z_{\tmin}(X)$, and consider the path amount of $b$ in the full path
$q(b,\B)$,
\begin{equation}
\label{TreeFunctionAndTotalAmountProof5}
\begin{aligned}
 e_{\B(X)}(b) & = \sum\limits_{b' \in \dot{q}(b,\B)} \big[\, m(b')-1\,
 \big] = \sum\limits_{b' \in \dot{q}(b, \B(a,X))} \big[\, m(b')-1\,
 \big] + m(X) - 1 \; = \\
 & = \; e_{\B(a,X)}(b) + m(X) - 1 \quad.
\end{aligned}
\end{equation}
As a consequence,
\begin{equation}
\label{TreeFunctionAndTotalAmountProof6}
 \sum\limits_{b \in z_{\tmax}(a)} n(b) \cdot e_{\B(a,X)}(b) =
 \sum\limits_{b \in z_{\tmax}(a)} n(b) \cdot e_{\B(X)}(b) - \big[\,
 m(X)-1\, \big] \cdot \# a \quad.
\end{equation}
If eqs.~(\ref{TreeFunctionAndTotalAmountProof4},
\ref{TreeFunctionAndTotalAmountProof6}) are inserted into the sum on
the RHS of eq.~(\ref{TreeFunctionAndTotalAmountProof3}) we obtain
\begin{equation}
\label{TreeFunctionAndTotalAmountProof7}
\begin{aligned}
 & \sum\limits_{a \in z_{\tmin}(X)}\; \sum\limits_{b \in \B(a,X)} n(b)
 \cdot \big[\, m(b)-1\, \big] = \\
 = - & \big[\, m(X)-1\, \big] \cdot \sum\limits_{a \in z_{\tmin}(X)}
 \# a + \sum\limits_{a \in z_{\tmin}(X)} \sum\limits_{b \in
 z_{\tmax}(a)} n(b) \cdot e_{\B(X)}(b) \quad.
\end{aligned}
\end{equation}
However,
\begin{equation}
\label{TreeFunctionAndTotalAmountProof8}
 \sum\limits_{a \in z_{\tmin}(X)} \# a = \# X = n(X) \quad,
\end{equation}
and
\begin{equation}
\label{TreeFunctionAndTotalAmountProof9}
 \sum\limits_{a \in z_{\tmin}(X)} \sum\limits_{b \in z_{\tmax}(a)}
 n(b) \cdot e_{\B(X)}(b) = \sum\limits_{b \in z_{\tmax}(X)} n(b) \cdot
 e_{\B(X)}(b) \quad.
\end{equation}
If eqs.~(\ref{TreeFunctionAndTotalAmountProof8},
\ref{TreeFunctionAndTotalAmountProof9}) are inserted into
eq.~(\ref{TreeFunctionAndTotalAmountProof7}) we obtain a contribution
$- [m(X)-1] n(X)$ which cancels the same term in
eq.~(\ref{TreeFunctionAndTotalAmountProof3}), so that $E_{\B(X)}$ on
the LHS of eq.~(\ref{TreeFunctionAndTotalAmountProof3}) is equal to
(\ref{TreeFunctionAndTotalAmountProof9}), which is what we have
claimed in eq.~(\ref{EqTreeFunctionAndTotalAmount1}). This finishes
the proof of (A).

Now we prove (B). If $\B(X)$ is complete, then $n(b)=1$ for all $b \in
z_{\tmax}(X)$. As a consequence,
eq.~(\ref{EqTreeFunctionAndTotalAmount1}) becomes
\begin{equation}
\label{TreeFunctionAndTotalAmountProof10}
 E_{\B(X)} = \sum_{b \in z_{\tmax}(X)} e_{\B(X)}(b) \quad,
\end{equation}
but, according to eq.~(\ref{TotalAmountOfTree}), the sum on the RHS of
(\ref{TreeFunctionAndTotalAmountProof10}) is just $G_{\B(X)}$ by
definition of $G$. This proves (B).
\end{proof}

\section{Minimal classes}  \label{MinimalClasses}

We now come to discuss the problem of minimizing the tree function
$E_{\B(X)}$ on certain sets of tree structures. We will need a couple
of new notions which we introduce in the sequel:

Consider the set $\M(X)$ of all tree structures $\B(X)$ over $X$. To
every $\B \in \M(X)$ we can uniquely assign the minimal partition
$z_{\tmin}(X)$ induced by $\B$ on $X$; this assignment will be denoted
by $z_{\tmin} : \M(X) \rightarrow \Z(X)$, $\B \mapsto
z_{\tmin}(X,\B)$. Given $z \in \Z(X)$, the inverse image
$z^{-1}_{\tmin}(z)$ is the set of all tree structures $\B$ over $X$
with the same minimal partition $z_{\min}(X)$ of $X$.

Let $n = \# X$. For $1 \le m \le n$, let $\M(X,m)$ denote the set of
all tree structures over $X$ whose minimal partition $z_{\tmin}(X)$
contains $m$ elements. Since all $\M(X,m)$ are disjoint, this defines
a partition of $\M(X)$,
\begin{equation}
\label{EqMinClasses1}
 \M(X) = \bigcup\limits_{1\le m \le n} \M(X,m) \quad.
\end{equation}

We recall that the tree function $E: \M(X) \rightarrow \NN$ sends
every tree over $X$ to the sum over all $n(b)\, [m(b)-1]$, as $b$
ranges through all nodes in the tree. We are interested in the minima
of this map, as $E$ is restricted to certain subsets of $\M(X)$. We
observe that it makes no sense to ask for the global minimum of $E$ on
$\M(X)$, as the answer is trivial: In this case the minimum clearly
is taken on the trivial tree $\B = \{X\}$, since $E_{\B} = G_{\B} =
0$. Meaningful results are obtained, however, if we first focus on the
subset of all {\it complete} trees $\C(X) \subset \M(X)$; this
inclusion is proper for $\# X \ge 2$. We write $\C(X,m)$ for the
set of all complete trees with $m$ elements in the minimal partition
of $X$. On the complete trees, the tree function $E$ coincides with
the total amount $G$, as follows from statement (B) in
theorem~\ref{TreeFunctionAndTotalAmount}. Now we define
\begin{equation}
\label{EqMinimalClasses2}
 \min(X) \equiv \min\limits_{\B \in \C(X)} E_{\B} \quad,
\end{equation}
and
\begin{equation}
\label{EqMinimalClasses3}
 \min(X,m) \equiv \min\limits_{\B \in \C(X, m)} E_{\B} \quad.
\end{equation}
In fact, $\min(X)$ is a function of $n = \# X$ only, and $\min(X,m)$
is a function of $n$ and $m$ only,
\begin{equation}
\label{EqMinimalClassesNotation1}
 \min(n) \equiv \min(X) \quad, \quad \min(n,m) \equiv \min(X,m) \quad.
\end{equation}
These minima exist, since all tree functions take their values in the
non-negative natural numbers. Thus it makes sense to speak of the set
of all complete trees
\begin{equation}
\label{EqMinimalClasses4}
 \Min(X) \equiv E^{-1}\big(\, \min(X)\, \big) \cap \C(X) \quad,
\end{equation}
on which the tree function $E$ actually takes its minimum. Similarly,
we introduce
\begin{equation}
\label{EqMinimalClasses5}
 \Min(X,m) \equiv E^{-1}\big(\, \min(X,m)\, \big) \cap \C(X,m)
 \quad.
\end{equation}
We term $\Min(X)$ the {\it global minimal class in
$\C(X)$}. $\Min(X,m)$ will be called the {\it minimal class in
$\C(X,m)$}.

\section{Divisions} \label{Divisions}

Given a natural number $n$, we can decompose $n$ into $m$ terms
according to $n= n_1 + \cdots + n_m$ with $1 \le m \le n$ in many
different ways, and for values of $m$ ranging from $1$ to $n$. For a
given $m$, the numbers $n_i$ can range between $1$ and $n$, and the
$n_i$ need not be mutually different. A decomposition of $n$ in this
form will be called a {\it division of $n$ into $m$ terms}. We can
regard it as an $m$-tupel $u= (n_1, \ldots, n_m)$ with positive
integer components, $n_i > 0$, such that $\sum n_i = m$. The set of
all divisions of $n$ into $m$ terms will be denoted by $U(n,m)$. If
$n$ is fixed and $m$ varies from $1$ to $n$, the collection of all
$U(n,m)$ defines a partition of the {\it set of all divisions $U(n)$
of $n$},
\begin{equation}
\label{DefDivisionsOfN}
 U(n) \equiv \bigcup\limits_{1\le m \le n} U(n,m) \quad.
\end{equation}
We introduce the {\it trivial division} $u_0 \equiv (n)$, and denote
the set of all nontrivial divisions of $n$ by $U^*(n) \equiv
U(n)-\{u_0\}$.

$U(n,m)$ is a proper subset of
\begin{equation}
\label{DefHDivision1}
 H(n,m) \equiv \Big\{\, h\in \RR^m\, \Big|\, \sum\limits_{i=1}^m h_i =
 n\, \Big\} \subset \RR^m \quad,
\end{equation}
which is a hyperplane in $\RR^m$ whose least Euclidean distance to the
origin is $\smallf{n}{\sqrt{m}}$. The element of $H(n,m)$ associated
with the least distance will be denoted by $\bar{h}$; it has
components $\bar{h} = ( \smallf{n}{m}, \ldots,
\smallf{n}{m})$. Usually, $n/m$ is not integer, so that $\bar{h} \neq
U(n,m)$. However, there are always elements $\bar{n}$ of $U(n,m)$ that
come closest to $\bar{h}$. The minimal distance between these elements
$\bar{n}$ and $\bar{h}$ ranges between $0$ and
$\smallf{\sqrt{m}}{2}$. If $\bar{h}$ coincides with a point in
$U(n,m)$, then $\bar{n} = \bar{h}$ is uniquely defined. The bigger the
distance between $\bar{h}$ and lattice points, the more elements
$\bar{n}$ there are: If $\bar{h}$ lies at the center of a cube formed
by elements of $U(n,m)$, then there are $2^m$ candidates for
$\bar{n}$, their distance from $\bar{h}$ being $\smallf{\sqrt{m}}{2}$
precisely. In this case each of the components $\bar{h}_i$ lies
exactly between two integer values, $\smallf{n}{m} \pm \smallf{1}{2}
\in \ZZ$; thus, $m$ must be even in this case. Whenever there is more
than one $\bar{n}$, i.e. more than one element of $U(n,m)$ with the
same minimal distance to $\bar{h}$, they must be related by
permutation of components.

There is another way to describe a division $n= n_1 + \cdots n_m$;
this is in terms of {\it occupation numbers} $t_k$ for all natural
numbers $k$ between $1$ and $n$ (and, in turn, even beyond), which
express how often $k$ appears as one of the terms $n_i$ in a given
decomposition of $n$. Obviously, the description of a division of $n$
into $m$ terms is determined by the set of occupation numbers $(t_1,
t_2, \ldots)$ uniquely up to permutation of the terms $n_i$ in the
sum. Here comes the detailed definition:

Let $n \in \NN$, let $1\le m \le n$. The $n$-tupel $t \equiv (t_1,
t_2, \ldots) \in \NN_0 \times \NN_0 \times \cdots$ will be called {\it
occupation numbers} of the division of $n$ into $m$ terms, if it
satisfies
\begin{subequations}
\label{DefOccupationNumbers}
\begin{align}
 \sum\limits_{k=1}^{n} t_k & = m \quad, \label{DefOccNum-a}
 \\
 \sum\limits_{k=1}^n k\cdot t_k & = n \quad. \label{DefOccNum-b}
\end{align}
\end{subequations}
The first sum says that the number of terms in the division of $n$ is
$m$; the second sum is just the decomposition of $n$. Clearly, for
$k>n$ all occupation numbers $t_k$ must vanish. For this reason we
will now focus on the finite sequences $t= (t_1, t_2, \ldots, t_n)$ of
occupation numbers rather than the infinite ones, so that $t$ ranges
in $\NN_0^n$.

The trivial division as expressed by occupation numbers is $t_0 \equiv
(0, \ldots, 0,1)$, i.e., $t_n=1$, and all other components
vanishing. The set of all occupation numbers of divisions of $n$ into
$m$ terms will be denoted by $T(n,m)$; the set of all occupation
numbers of divisions of $n$ will be written as $T(n)$. The occupation
numbers of nontrivial divisions comprise the set $T^*(n)$. Clearly,
$t_n=0$ for every nontrivial $t \in T^*(n)$.

The relation between divisions $u$ and their associated occupation
numbers $t$ is as follows: Every division $u= (n_1, \ldots, n_m)$
defines a unique $n$-tupel of occupation numbers $\kappa(u) \equiv
(t_1, \ldots, t_n)$ by
\begin{equation}
\label{OccupNum2}
 \kappa(u)_a = t_a \equiv \sum\limits_{i=1}^m \delta_{a, n_i} \quad.
\end{equation}
It follows readily that this indeed satisfies
(\ref{DefOccupationNumbers}). Furthermore, every $n$-tupel $t$ of
occupation numbers defines a unique naturally ordered division $u$ of
$n$ by $m$ terms, $u_1 \le u_2 \le \cdots \le u_m$. Now the inverse
image $\kappa^{-1}(t)$ of an occupation number tupel $t$ is just the
set of all divisions $u'$ that are related to the naturally-ordered
division $u$ by permutation of components. Thus, every such inverse
image has a naturally-ordered representative. We conclude that there
is a 1--1 relation between naturally-ordered divisions of $n$ and
occupation numbers.

\section{Partitions and divisions} \label{PartitionsAndDivisions}

Let $n= \# X$, let $z$ be an arbitrary partition of $X$, not
necessarily related to a tree structure over $X$. Assume that the
partition $z$ contains $m$ elements, $m= \# z$, where $z= \{b_1,
\ldots, b_m\}$. $z$ defines a division $u(z)$ of $n$ into $m$ terms by
$u = (\# b_1, \ldots, \# b_m)$. This defines the {\it $u$-map} $u :
\Z(X) \rightarrow U(n)$, $z \mapsto u(z)$. The associated occupation
number will be written as $t(z)$ and has components
\begin{equation}
\label{PartAndDiv1}
 t(z)_a \equiv \sum\limits_{b \in z} \delta_{a, \# b} \quad
\end{equation}
for $a= 1, \ldots, n$. $t(z)_a$ will be called the {\it $a$-th
occupation number of the partition $z$}. This defines the {\it
$t$-map} $t: \Z(X) \rightarrow T(n)$, $z \mapsto t(z)$; it sends every
partition of $X$ to the associated $n$-tupel of occupation
numbers. The $u$-, $t$-maps are obviously surjective, since for every
division of $n$ into $m$ terms one can construct an associated
partition of $X$.

From the surjectivity of $u$ and $t$ and the fact that the map
$z_{\tmin}$ sends $\M(X)$ onto the set of all partitions $\Z(X)$ we
find $U(n) = (u \circ z_{\tmin})(\M(X))$ and $T(n) = (t \circ
z_{\tmin})(\M(X))$, and furthermore, $U(n,m) = (u \circ
z_{\tmin})(\M(X,m))$ and $T(n,m) = (t \circ z_{\tmin})(\M(X,m))$.

The distinct occupation number $t_{\tmin}(X) \equiv (t \circ
z_{\tmin})(\B)$ will be called the {\it minimal division of $n= \# X$}
in $\B(X)$.

\begin{definition}[Integer quotient] \label{IntegerQuotient}
For $n\in \NN_0$, $m \in \NN$, let
\begin{equation}
\label{EqIntegerQuotient}
 \left[ \frac{n}{m} \right] \equiv \big\{\, n' \in \NN_0\, \big|\, n'
 \cdot m \le n\, \big\}
\end{equation}
denote the {\it integer quotient of $n$ by $m$}.
\end{definition}

\section{Optimal division} \label{OptimalDivision}

Let $1 \le m \le n$. Let $\nu \equiv \left[\smallf{n}{m} \right]$ be
the integer quotient of $n$ by $m$; then $n = \nu \cdot m + r$ with $0
\le r < m$. We construct a division of $n$ into $m$ terms according to
\begin{equation}
\label{EqOptDiv1}
 ( \nu, \ldots, \nu, \nu+1, \ldots, \nu+1) \quad,
\end{equation}
with $(m-r)$ occurrences of $\nu$ and $r$ occurrences of
$(\nu+1)$. The associated occupation number is denoted as $\bar{t}
\equiv \bar{t}(n,m) = (\bar{t}_1, \ldots, \bar{t}_n)$, with
$\bar{t}_{\nu} = m-r$, $\bar{t}_{\nu+1} = r$, and $\bar{t}_{\lambda} =
0$ for $\lambda \not\in \{\nu, \nu+1\}$. Consider the inverse image
$\kappa^{-1}(\bar{t})$ of $\bar{t}$ under $\kappa$; every
representative of this set will be called {\it optimal division of $n$
by $m$}, and will be denoted by $\bar{n}$. Obviously, the optimal
divisions come closest to the $m$-tupel $\bar{h} = (\smallf{n}{m},
\ldots, \smallf{n}{m}) \in H(n,m) \subset \RR^m$, where $\bar{h}$ is
the element in $H(n,m)$ with least Euclidean distance to the origin;
thus, they coincide with the objects $\bar{n}$ introduced in
section~\ref{Divisions}. We observe that $\kappa^{-1}(\bar{t})$ is the
set of all elements $\bar{n}$ of $U(n,m)$ for which the Euclidean norm
\begin{equation}
\label{EqOptDiv2}
 \left\| \bar{n} - \bar{h} \right\| \le \frac{\sqrt{m}}{2} \quad.
\end{equation}

We now prove an important lemma about optimal divisions:
\begin{lemma}[Optimal divisions] \label{LemmaOptimalDivisions}
Let $\|u\| = \sqrt{ \sum_{i=1}^m u_i^2}$ denote the Euclidean norm of
an element $u \in \RR^m$. Let $u= (n_1, \ldots, n_m)$ be an element of
$U(n,m)$. Then there exists a finite sequence $u^0, u^1, \ldots, u^f$
of elements in $U(n,m)$ with $u^0 = u$, $u^f = \bar{n}$ for some
$\bar{n} = \kappa^{-1}(\bar{t})$, such that
\begin{equation}
\label{EqLemOptDiv}
 \left\| u^0\right\| > \left\| u^1\right\| > \cdots > \left\| u^f
 \right\| \quad,
\end{equation}
and the step $u^{\alpha} \rightarrow u^{\alpha+1}$ involves alteration
of two components of $u^{\alpha}$ only.
\end{lemma}
\begin{proof}
Denote $M \equiv \{1, \ldots, m\}$ for short. For $(i,j) \in M^2$,
$i\neq j$, we define an operation $S_{ij}$ on elements $h \in \RR^m$
by
\begin{equation}
\label{EqLemOptDivProof1}
 S_{ij}(h_1, \ldots, h_m) \equiv (h_1, \ldots, h_i + 1, \ldots, h_j -
 1, \ldots, h_m) \quad,
\end{equation}
i.e., all components except $h_i$ and $h_j$ remain the same. By
construction, $S_{ij}$ preserves $H(n,m)$, for if $h \in H(n,m)$ then
so is $S_{ij} h$.

--- We prove the statement: Let $u \in U(n,m)$, let $\Delta \equiv u -
\bar{h}$. If $\Delta_j - \Delta_i \le 1$ for all $i,j \in M^2$, then
\begin{equation}
\label{EqLemOptDivProof2}
 u = \bar{n} \in \kappa^{-1}(\bar{t}) \quad.
\end{equation}

{\it Proof of (\ref{EqLemOptDivProof2}):} If $\Delta = 0$ then the
statement is trivial; hence assume $\Delta \neq 0$. Let
$\Delta_{\tmax}$ denote the maximal element in $\{\Delta_1, \ldots,
\Delta_m\}$. $\Delta_{\tmax}$ is certainly $> 0$; for, $\sum \Delta_i
= \sum u_i - \sum \bar{h}_i = 0$, and there must be nonzero components
of $\Delta_i$. Our starting assumption says $\Delta_{\tmax} - \Delta_i
\le 1$, but on the other hand, $\Delta_i \le \Delta_{\tmax}$, hence
\begin{equation}
\label{EqLemOptDivProof3}
 0 \le \Delta_{\tmax} - \Delta_i \le 1 \quad \text{for all $i$} \quad.
\end{equation}
However, $\Delta_i - \Delta_j = u_i - u_j \in \ZZ$, hence the same
must be true for the quantities $\Delta_{\tmax} - \Delta_i$. We
conclude that $\Delta_{\tmax} - \Delta_i \in \{0, 1\}$. We have
altogether $m$ components $\Delta_i$, which can take values of either
$\Delta_{\tmax}$ or $\Delta_{\tmax}-1$. Suppose there are $(m-r)$
components $\Delta_i = \Delta_{\tmax}$, and $r$ components of the form
$\Delta_{\tmax}-1$, where $0 \le r \le m$. We cannot have $r=m$, for
otherwise none of the $\Delta_i$ would take the maximal value
$\Delta_{\tmax}$; thus, $r<m$. The sum over all $\Delta_i$ must
vanish, from which it follows that $m \Delta_{\tmax} = r$. An easy
computation now gives $\|\Delta\|^2 = \smallf{r(m-r)}{m}$. If $m$ is
fixed, the expression on the RHS is zero for $r=0$ and becomes maximal
for $r= \smallf{m}{2}$, in which case it takes the value
$\smallf{m}{4}$. Hence $\| u - \bar{h} \| \le \smallf{\sqrt{m}}{2}$,
which implies that $u= \bar{n}$ by eq.~(\ref{EqOptDiv2}). This proves
the statement~(\ref{EqLemOptDivProof2}).

--- Now we prove our lemma: We describe step 1 in constructing the
series~(\ref{EqLemOptDiv}): Let $\Delta^0 \equiv u^0 - \bar{h}$. If
$u^0 = \bar{n}$, there is nothing to prove. If $u^0 \neq \bar{n}$, we
conclude from statement~(\ref{EqLemOptDivProof2}) that there exists a
pair $(i,j) \in M^2$ with $i \neq j$ such that $\Delta^0_j -
\Delta^0_i > 1$; since the left-hand side must be integer we must
have, in fact, that $\Delta^0_j - \Delta^0_i \ge 2$. Now we define the
new element $u^1 \equiv S_{ij} u^0$ for this choice of $(i,j)$. Let
$\Delta^1 \equiv u^1 - \bar{h} = S_{ij} \Delta^0$. Since $\bar{h}$ has
least distance to the origin, it is perpendicular to the hyperplane
$H(n,m)$, whereas $\Delta^0, \Delta^1$ lie in this plane. Hence, by
Pythagoras,
\begin{equation}
\label{EqLemOptDivProof4}
\begin{aligned}
 \left\| u^0 \right\|^2 & = \left\| \bar{h} \right\|^2 + \left\|
 \Delta^0 \right\|^2 \quad, \\
 \left\| u^1 \right\|^2 & = \left\| \bar{h} \right\|^2 + \left\|
 \Delta^1 \right\|^2 \quad,
\end{aligned}
\end{equation}
or $\|u^1 \|^2 - \|u^0 \|^2 = \| S_{ij} \Delta^0 \|^2 - \|\Delta^0
\|^2$. The last expression is just $2 ( \Delta^0_i - \Delta^0_j + 1)$,
which must be $\le -2$ owing to $\Delta^0_j - \Delta^0_i \ge 2$. Thus,
\begin{equation}
\label{EqLemOptDivProof5}
 \left\| u^1\right\|^2 \le \left\| u^0 \right\|^2 - 2 < \left\| u^0
 \right\|^2 \quad,
\end{equation}
and only two components of $u^0$, namely $u^0_i$ and $u^0_j$ have been
altered. This finishes step 1. In step 2 we check whether $u^1 =
\bar{n}$ for some $\bar{n}$; if yes, the process terminates; if no, it
continues in the same manner. Since every step $\alpha$ involves a
decrease of $\|\Delta^{\alpha} \|^2$ by at least $-2$, the process
must terminate after a finite number of steps.
\end{proof}

\section{Optimal trees} \label{OptimalTrees}

A tree structure $\B_o = \B_o(X)$ over the set $X$ is called {\it
optimal over $X$}, if $\B_o$ is complete, and
\begin{equation}
\label{DefOptimalTree}
 t\big(\, z_{\tmin}(b)\, \big) = \bar{t}\big(\, n(b), 2\, \big)
\end{equation}
for all non-terminal elements $b \in \B_o$. This means that every node
$b$ not belonging to the maximal partition $z_{\tmax}(X)$ of $X$ is
partitioned into two halves which are as close to being equal as
possible, when stepping to the next level in the tree; and every
terminal node contains only one element. The set of all optimal trees
over $X$ forms (for $\# X> 2)$ a proper subset of $\C(X)$, which will
be denoted by $\O(X)$.

\section{Minimal classes in $T(n,m)$} \label{MinimalClassesTNM}

Every minimal tree $\B \in \Min(X)$ maps into a certain partition $z$
under $z_{\tmin}$, and into a certain occupation number $t$ under the
$t$-map. We shall be interested in the image of $\Min(X)$ under this
sequence of maps, which we will denote as
\begin{equation}
\label{DefMinClasses1}
 T_{\tmin}(n) \equiv \left( t \circ z_{\tmin} \right)\big(\, \Min(X)\,
 \big) \quad,
\end{equation}
and which we shall call the {\it global minimal class in $T(n)$}. For
$1 \le m \le n$, the set
\begin{equation}
\label{DefMinClasses2}
 T_{\tmin}(n,m) \equiv \left( t \circ z_{\tmin} \right)\big(\,
 \Min(X,m)\, \big)
\end{equation}
will be termed the {\it minimal class in $T(n,m)$}.

We note that we now have several distinct classes of occupation
numbers in $T(n)$: We have the class containing all optimal divisions
of $n$ by $m$, $\{\, \bar{t}(n,1), \bar{t}(n,2),$ $\ldots,$
$\bar{t}(n,n)\, \}$; and on the other hand, the classes
$T_{\tmin}(n,m)$. The relation between these will be investigated in
the following developments.

Now let $t \in T(n)$ arbitrary. We can study its inverse image
$\left(t \circ z_{\tmin} \right)^{-1}(t) \cap \C(X)$ in
$\C(X)$. To every tree in this set we can assign the associated tree
function $E$; thus it makes sense to ask on which trees $\B \in
\left(t \circ z_{\tmin} \right)^{-1}(t) \cap \C(X)$, for a given
division $t$, the tree function $E$ assumes its minimum. This minimum
will be denoted by $\min(t)$; hence
\begin{equation}
\label{DefMinClasses3}
 \min(t) \equiv \min\limits_{\B \in \left(t \circ
 z_{\tmin}\right)^{-1}(t) \cap \C(X)} E_{\B} \quad.
\end{equation}
The associated subset of trees in $\left( t \circ
z_{\tmin}\right)^{-1}(t) \cap \C(X)$ that actually take this minimum
will be denoted as $\Min(t)$,
\begin{equation}
\label{DefMinClasses4}
 \Min(t) \equiv E^{-1}\big(\, \min(t)\, \big) \cap \left( t\circ
 z_{\tmin}\right)^{-1}(t) \cap \C(X) \quad.
\end{equation}

\section{Bases and integer logarithm} \label{BasesIntegerLogarithm}

Let $L \in \NN$ with $L \ge 2$. Then the set
\begin{equation}
\label{DefBasis}
 \BB_L \equiv \big\{\, L^k \,|\, k \in \NN_0\, \}
\end{equation}
will be called {\it basis over $L$}. The set $\BB_2$ we shall also
call {\it binary basis}. If no confusion is likely, $\BB_2$ will be
simply denoted by $\BB$.

\begin{definition}[Integer logarithm] \label{DefIntegerLogarithm}
Let $n \in \NN$ be a natural number. The {\em integer logarithm
$\lg_L(n)$} of $n$ with respect to $L$ is defined as
\begin{equation}
\label{DefIntegerLog}
 \lg_L(n) \equiv \max \Big\{\, k \in \NN_0\, \Big|\, L^{k} \le n\,
 \Big\} \quad.
\end{equation}
\end{definition}
If no confusion is likely, the integer logarithm of $n$ with respect
to $2$ will simply be written $\lg(n) \equiv \lg_2(n)$. Clearly,
$\lg_L$ is a monotonically increasing function on $\NN$.

\begin{proposition}[Properties of integer logarithm]
\label{PropertiesIntegerLogarithm}
The integer logarithm satisfies the following inequalities:
\begin{subequations}
\label{EqPropIntLog}
\begin{enumerate}
\item Let $n,n' \in \NN$ with $n' \ge n$. Then
\begin{equation}
\label{EqPropIntLog1}
 \lg_L(n') \ge \lg_L(n) \quad.
\end{equation}
\item Let $n,n' \in \NN$. Then
\begin{equation}
\label{EqPropIntLog2}
 \lg_L(n) + \lg_L(n') \le \lg_L(n \cdot n') \quad.
\end{equation}
\item Let $p \in \NN_0$. Then
\begin{equation}
\label{EqPropIntLog3}
  p \cdot \lg_L(n) \le \lg_L\left(n^p\right) \quad.
\end{equation}
\end{enumerate}
\end{subequations}
\end{proposition}
\begin{proof}
$\lg_L(n)$ is the maximum in the set of those integers $k$ which
satisfy $L^k \le n$. As a consequence, $L^{\lg_L(n)} \le n'$. Hence
$\lg_L(n)$ lies in the set of those integers $k'$ which satisfy
$L^{k'} \le n'$ and consequently must be less than or equal to its
maximum. This proves (\ref{EqPropIntLog1}).

 Let $k= \lg_L(n)$ and $k' = \lg_L(n')$. Then $L^k \le n$ and $L^{k'}
\le n'$, from which it follows that $L^{k+k'} \le n n'$, hence $k+k'
\le \lg_L{n n'}$. The converse is not necessarily true. ---
Furthermore, from $L^k \le n$ it follows that $L^{p k} \le n^p$, hence
$pk \le \lg_L(n^p)$.
\end{proof}
\begin{lemma}[Standard decomposition] \label{StandardDecomposition}
Every natural number $n \in \NN_{\ge 0}$ has a standard decomposition
\begin{subequations}
\label{EqStandardDec}
\begin{align}
 n & = 2^{\lg(n)} + R \quad, \label{ESDa} \\
 R & < \frac{n}{2}\, , \; 2^{\lg(n)} \quad. \label{ESDb}
\end{align}
\end{subequations}
\end{lemma}
\begin{proof}
We show that $R$ must indeed be limited to be $< n/2$. Suppose to the
contrary, then $2R \ge n$ for some $n$. It follows that $2n =
2^{\lg(n)+1} + 2R \ge 2^{\lg(n)+1} + n$ or $n \ge 2^{\lg(n)+1}$,
implying $\lg(n) \ge \lg(n)+1$, which is a contradiction. Hence
(\ref{ESDa}) holds. -- Similarly, if $R \ge 2^{\lg(n)}$, then $n \ge
2^{\lg(n)+1}$, leading to a contradiction as before; thus,
(\ref{ESDb}) is true.
\end{proof}
\begin{lemma} \label{EvenIntegers}
The equation
\begin{equation}
\label{EqEvenIntegers}
 \lg(2\nu + 1) = \lg(2 \nu) = \lg(\nu) + 1
\end{equation}
holds for all integers $\nu \in \NN$.
\end{lemma}
\begin{proof}
We use the decomposition (\ref{ESDa}) for $2\nu$,
\begin{equation}
\label{EqEvenIntProof2}
 2\nu = 2^{\lg(2\nu)} + R \quad, \quad 0 \le R < 2^{\lg(2\nu)}\, ,\;
 \nu \quad.
\end{equation}
Assume that the first equation in (\ref{EqEvenIntegers}) is not true,
but rather
\begin{equation}
\label{EqEvenIntProof3}
 \lg(2\nu+1) = \lg(2\nu) + 1 \quad.
\end{equation}
We then have a chain of inequalities
\begin{equation}
\label{EqEvenIntProof4}
\begin{aligned}
 2\nu+1 & \ge 2^{\lg(2\nu+1)} = 2\cdot 2^{\lg(2\nu)} = \\
 & = 2 \cdot \left[ 2^{\lg(2\nu)} + R - R \right] = 2\cdot \left[
 2\nu-R \right] \quad.
\end{aligned}
\end{equation}
It follows that
\begin{equation}
\label{EqEvenIntProof5}
 R \ge \nu - \frac{1}{2} \quad,
\end{equation}
but $R, \nu$ are integers, and therefore we must have
\begin{equation}
\label{EqEvenIntProof6}
 R \ge \nu \quad,
\end{equation}
which contradicts the second inequality on the RHS of
(\ref{EqEvenIntProof2}).

We now prove the second equation in (\ref{EqEvenIntegers}): Let $k
\equiv \lg(\nu)$. Then $2^k \le \nu$, but $2^{k+1} > \nu$. By
multiplying these inequalities with a factor of $2$ we find $2^{k+1}
\le 2\nu$, but $2^{k+2} > 2\nu$. It follows that $k+1$ has the maximum
property with respect to $2\nu$, as required in
definition~\ref{DefIntegerLogarithm}, and therefore $\lg(2\nu) = k+1$,
which proves the second equation in (\ref{EqEvenIntegers}).
\end{proof}

\section{Optimal amount} \label{OptimalAmount}

\begin{theorem}[Amount of optimal trees] \label{AmountOptimalTrees}
Let $n = \# X$ and $\B_o \in \O(X)$. Let $\lg(n)$ denote the integer
logarithm with respect to $2$. Then
\begin{equation}
\label{EqOptAmount}
 E_{\B_o} = G_{\B_o} = n\cdot \lg(n) + 2 \cdot \big[ n - 2^{\lg(n)}
 \big] \quad.
\end{equation}
This value is the same for all $\B_o \in \O(X)$ and depends only on
$n$.
\end{theorem}
\begin{definition} \label{DefValueOptAmount}
The common value of the amount of the optimal trees will be denoted by
\begin{equation}
\label{EqDefValueOptAmount}
 E(n) = G(n) \equiv E_{\B_o} = G_{\B_o} \quad.
\end{equation}
\end{definition}
\begin{proof}
By induction with respect to $n$. The statement is clear for $n=1$,
since in this case, $E_{\B_o} = G_{\B_o} = 0$, and $\lg(1)= 0$.

Now perform the induction: Assume that eq.~(\ref{EqOptAmount}) holds
for all $1 \le n' < n$. Let $X$ be a set with $\# X = n$, let $\B_o \in
\O(X)$. Then the minimal partition $z_{\tmin}(X)= \{b_1, b_2\}$ of
$X$ has two elements $b_1$ and $b_2$ whose cardinalities are as close
to $n/2$ as possible. Use formula~(\ref{EqSplittingLemmaCorollary2}),
together with the fact that $m(X) = 2$,
\begin{equation}
\label{EqOptAmProof1}
 G_{\B_o} = n + \sum\limits_{b \in z_{\min}(X)} G_{\O(b,X)} \quad.
\end{equation}
We have $\# b \le n-1$ for all $b \in z_{\tmin}(X)$, and hence
\begin{equation}
\label{EqOptAmProof2}
 G_{\O(b,X)} = \#b \cdot \lg(\#b) + 2 \cdot \left[ \#b- 2^{\lg(\#b)}
 \right]
\end{equation}
by assumption. We now must distinguish whether $n$ is even or odd:

\underline{Case 1:} \quad $n=2\nu + 1$. We apply
formula~(\ref{EqOptAmProof1}), using the fact that $\#b_1 = \nu+1$ and
$\#b_2 = \nu$. This gives
\begin{equation}
\label{EqOptAmProof3}
 G_{\B_o} = 3n + (\nu+1) \cdot \lg(\nu+1) + \nu \cdot \lg(\nu) -
 2^{\lg(\nu+1) +1} - 2^{\lg(\nu) +1} \quad.
\end{equation}
Two subcases must be considered: $\lg(\nu) = \lg(\nu+1)$, or $\lg(\nu)
+ 1 = \lg(\nu+1)$. Consider first the case $\lg(\nu) =
\lg(\nu+1)$,
\begin{equation}
\label{EqOptProof4}
 G_{\B_o} = 3n + n \cdot \lg(\nu) - 2^{\lg(\nu)+2} \quad.
\end{equation}
The first two terms give $2n + n\cdot \lg(n)$, upon using
eq.~(\ref{EqEvenIntegers}) in lemma~\ref{EvenIntegers}. Multiple
applications of the same equation then produce
eq.~(\ref{EqOptAmount}). Now consider subcase $\lg(\nu+1) = \lg(\nu) +
1$: This is the case if and only if $\nu = 2^K -1$, where $K \in
\NN_{>0}$. Therefore, $K= \lg(\nu+1)$. On using $\lg(n) = \lg(\nu)+1$
we find
\begin{equation}
\label{EqOptProof5}
 G_{\B_o} = 2n + n \lg(n) + \nu + 1 - 2^K - 2^{K+1} \quad.
\end{equation}
But $\nu+1- 2^K =0$, and hence we arrive at (\ref{EqOptAmount}) again.

\underline{Case 2:} \quad $n = 2\nu$. In this case, $\#b_1 = \#b_2 =
\nu$, and formula~(\ref{EqOptAmProof1}) gives
\begin{equation}
\label{EqOptProof6}
 G_{\B_o} = n + 2\nu \left[ \lg(\nu)+1 \right] + 2\nu - 2 \cdot
 2^{\lg(\nu)} \quad,
\end{equation}
which gives again (\ref{EqOptAmount}), on using $\lg(n) = \lg(\nu)+1$.
\end{proof}

\begin{lemma}[Monotonicity of optimal amount]
 \label{MonotonicityOptimalAmount}
The optimal amount is a monotonically increasing function of $n$. In
particular,
\begin{equation}
\label{EqLemMonOptAm}
 G(n+1) - G(n) = \lg(n) + 2 \quad \text{for all $n \in \NN$} \quad.
\end{equation}
\end{lemma}
\begin{proof}
 The result follows directly from eq.~(\ref{EqOptAmount}) for both
 cases $\lg(n) = \lg(n+1)$ and $\lg(n) + 1 = \lg(n+1)$.
\end{proof}

Using this lemma we can prove:
\begin{theorem}[Total amount for unsymmetric divisions]
\label{AmountUnsymmetricDivisions}
Let $n \in \NN_{>0}$. Consider two divisions of $n$ into two terms,
\begin{subequations}
\begin{equation}
\label{EqAmUnDiv1}
 n_1 + n_2 = n'_1 + n'_2 \quad.
\end{equation}
 If
\begin{equation}
 \left( n'_1\right)^2 + \left( n'_2 \right)^2 \ge n_1^2 + n_2^2 \quad,
 \label{EqAmUnDiv2}
\end{equation}
then
\begin{equation}
  G(n'_1) + G(n'_2) \ge G(n_1) + G(n_2) \quad. \label{EqAmUnDiv3}
\end{equation}
\end{subequations}
\end{theorem}
\begin{proof}
Eq.~(\ref{EqAmUnDiv1}) implies that some $\Delta$ exists such that
$n'_1 = n_1 + \Delta$ and $n'_2 = n_2 - \Delta$. Without loss of
generality we can assume that $\Delta \ge 0$. If $\Delta =0$ then
there is nothing to prove, hence we can assume $\Delta> 0$. Then
(\ref{EqAmUnDiv3}) is equivalent to
\begin{equation}
\label{EqAmUnProof1}
 G(n_1 + \Delta) - G(n_1) \ge G(n_2) - G(n_2 - \Delta) \quad.
\end{equation}
The LHS and RHS can be written as sums over differences $G(n_1+\Delta)
- G(n_1+\Delta -1)$, etc. Using eq.~(\ref{EqLemMonOptAm}) we find that
(\ref{EqAmUnProof1}) is equivalent to
\begin{equation}
\label{EqAmUnProof2}
 \sum\limits_{j=0}^{\Delta-1} \lg(n_1+j) \ge
 \sum\limits_{j=0}^{\Delta-1} \lg(n_2 - \Delta + j) \quad.
\end{equation}
Now a short computation shows that (\ref{EqAmUnDiv2}) implies
\begin{equation}
\label{EqAmUnProof3}
 n_1 \ge n_2 - \Delta \quad,
\end{equation}
Now eq.~(\ref{EqPropIntLog1}) in
proposition~\ref{PropertiesIntegerLogarithm} shows that $\lg(n_1+j)
\ge \lg(n_2- \Delta +j)$ for each pair of terms in
(\ref{EqAmUnProof2}). Thus (\ref{EqAmUnDiv3}) follows.
\end{proof}

This theorem is used in proving the important
\begin{theorem}[Amount of non-optimal divisions]
\label{AmountNonOptimalDivisions}
Let $u= (n_1, \ldots, n_m) \in U(n,m)$, let $\bar{n} = (\bar{n}_1,
\ldots, \bar{n}_m)$ $\in \kappa^{-1}(\bar{t})$ be an optimal division
of $n$ by $m$. Then
\begin{equation}
\label{EqAmNonOptDiv}
 \sum\limits_{i=1}^{m} G(n_i) \ge \sum\limits_{i=1}^m G(\bar{n}_i)
 \quad.
\end{equation}
\end{theorem}
\begin{proof}
Clearly, the RHS of (\ref{EqAmNonOptDiv}) is independent of the
representative $\bar{n} \in \kappa^{-1}(\bar{t})$, as the
representatives differ only by permutations of components. According
to lemma~\ref{LemmaOptimalDivisions} there exists a finite sequence
$u^0, u^1, \ldots, u^f$ of elements in $U(n,m)$ with $u^0 = u$ and
$u^f = \bar{n}$ for some $\bar{n} \in \kappa^{-1}(\bar{t})$, such that
$\|u^0\| > \|u^1 \| > \cdots > \|u^f \|$, and the step $u^{\alpha}
\rightarrow u^{\alpha+1}$ involves alteration of two components of
$u^{\alpha}$ only, namely $u^{\alpha+1}_i = u^{\alpha}_i + 1$ and
$u^{\alpha+1}_j = u^{\alpha}_j - 1$. As a consequence,
\begin{equation}
\label{EqAmNonOptDivProof1}
 \left( u^{\alpha}_i \right)^2 + \left( u^{\alpha}_j \right)^2 >
 \left( u^{\alpha+1}_i \right)^2 + \left( u^{\alpha+1}_j \right)^2
 \quad,
\end{equation}
whereas $u^{\alpha}_k = u^{\alpha+1}_k$ for all $k \not\in
\{i,j\}$. Thus, using theorem~\ref{AmountUnsymmetricDivisions} with
(\ref{EqAmNonOptDivProof1}) implies that
\begin{equation}
 \sum_{i=1}^m G(u^{\alpha}_i) \ge \sum_{i=1}^m G(u^{\alpha+1}_i)
 \quad.
\end{equation}
This inequality holds for every step $\alpha \rightarrow \alpha+1$,
and hence (\ref{EqAmNonOptDiv}) follows.
\end{proof}

\section{Preoptimized trees} \label{PreoptimizedTrees}

The concept of preoptimization is required as a necessary intermediate
step in order to solve the problem of finding the global minimal class
$\Min(n)$. Let $n = \# X$.
\begin{definition}[Preoptimized trees] \label{DefPreoptimizedTrees}
A tree $\B \in \M(X)$ is called {\em preoptimized} if every subtree
$\B(b,X)$ of $\B$ based on elements $b \in z_{\tmin}(X)$ in the
minimal partition of $X$ in $\B$ is optimal.
\end{definition}
Thus, the only "degrees of freedom" of varying a preoptimized tree are
the different choices of minimal partitions $z_{\tmin}(X)$, where
these choices can be effectively described by the set of all divisions
$U(n)$ of $n$ into $m$ terms, for $m= 1, \ldots, n$. Every
preoptimized tree is complete by definition. The subset of all
preoptimized trees over $X$ in $\M(X)$ will be denoted by $p\O(X)
\subset \C(X)$. $p\O(X)$ is comprised of the disjoint subsets
$p\O(X,m)$ of preoptimized trees with $m$ elements in the minimal
partition $z_{\tmin}(X)$; hence we have a partition of $p\O(X)$
according to $p\O(X) = \bigcup\limits_{1 \le m \le n}
p\O(X,m)$. Furthermore, we define
\begin{equation}
\label{DefPreOptStar}
 p\O^*(X) = \bigcup\limits_{2\le m\le n} p\O(X,m) = p\O(X) - \{\{X\}\}
\quad.
\end{equation}
The structure of the subsets $\O(X), p\O(X), p\O^*(X), \ldots$, etc.,
is independent of the nature of the underlying set $X$ but depends
only on the number $n = \# X$ of elements in it. We can therefore
write $\O(X) = \O(n)$, $p\O(X) = p\O(n)$, $p\O^*(X) = p\O^*(n)$, etc.,
when appropriate.

On the subsets just described, the tree function $E$ coincides with
the total amount by theorem~\ref{TreeFunctionAndTotalAmount}, since
all trees are complete. There, $E = G$ takes the minima
\begin{subequations}
\label{DefPreOptMin}
\begin{equation}
 p \min(X) \equiv \min\limits_{\B \in p\O^*(X)} E_{\B} \quad,
 \label{DefPreOptMin1}
\end{equation}
and
\begin{equation}
 p \min(X,m) \equiv \min\limits_{\B \in p\O(X,m)} E_{\B}
 \quad. \label{DefPreOptMin2}
\end{equation}
\end{subequations}
In (\ref{DefPreOptMin1}) we have restricted the trees to the set
$p\O^*(X)$, since for the trivial preoptimized partition $z_{\tmin}(X)
= \{X\}$, there is nothing to optimize since there are no subtrees;
and the total amount $G$ of this tree, as well as the corresponding
tree function $E$, is zero.

In accord with definitions (\ref{DefPreOptMin}) we introduce the sets
of all preoptimized trees for which $G = E$ takes the corresponding
minima:
\begin{equation}
\label{DefPreOptMinSets}
\begin{aligned}
 \Min_p(X) & \equiv \Min_p(n) \equiv E^{-1}\big(\, p \min(X)\, \big)
 \cap p\O^*(X) \quad, \\
 \Min_p(X,m) & \equiv \Min_p(n,m) \equiv E^{-1}\big(\, p \min(X,m)\,
 \big) \cap p\O(X,m) \quad.
\end{aligned}
\end{equation}
Obviously, $(t\circ z_{\tmin})(p\O^*(X)) = T^*(n)$, and $(t\circ
z_{\tmin})(p\O(X,m)) = T(n,m)$.  Hence $p\O(X,m)$ can be partitioned
according to
\begin{equation}
\label{PartPreOptSet}
 p\O(X,m) = \bigcup\limits_{t \in T(n,m)} \Big[\, (t\circ
 z_{\tmin})^{-1}(t) \cap p\O(X)\, \Big] \quad.
\end{equation}

In all of the discussions so far the nature of the set $X$ was
immaterial; the only thing that matters is the number $n= \# X$ of
elements in $X$. Thus, we could replace in each of the quantities
above the symbol $X$ by $n$.

Now let $t \in T(n)$, $m = \sum_{i=1}^n t_i$ and $\B \in (t\circ
z_{\tmin})^{-1}(t) \cap p\O(X)$ such that $m = \# z_{\tmin}(X)$ for
the corresponding minimal partition of $X$. From
eq.~(\ref{EqSplittingLemmaCorollary2}) in
corollary~\ref{SplittingLemmaCorollary} we have
\begin{equation}
\label{EqPreOptTrees1}
 E_{\B} = G_{\B} = n(m-1) + \sum_{b \in z_{\tmin}(X)} G_{\B(b,X)}
 \quad.
\end{equation}
But all subtrees $\B(b,X)$ are optimal, hence $G_{\B(b,X)}$ coincides
with $G(\# b)$ according to eq.~(\ref{EqOptAmount}) in
theorem~\ref{AmountOptimalTrees}, and the first term $n(m-1)$ is
constant for fixed $t$. Here, the common value of all optimal trees
with $\# b$ elements in the underlying set is denoted as $G(\#b)$,
according to definition~\ref{DefValueOptAmount}. Thus $E_{\B}$ is
constant on $(t \circ z_{\tmin})^{-1}(t) \cap p\O(X)$ and hence {\it
descends} to a map, again denoted by
\begin{equation}
\label{descend1}
 E: T(n) \rightarrow \NN \quad, \quad E(t) \equiv E_{\B} \quad,
\end{equation}
for any choice of representative $\B \in (t \circ z_{\tmin})^{-1}(t)
\cap p\O(X)$. Now (\ref{EqPreOptTrees1}) can be expressed as
\begin{equation}
\label{EqPreOptTrees2}
 E(t) = n(m-1) + \sum_{k=1}^n t_k \cdot G(k) \quad,
\end{equation}
for all $t \in T(n,m)$. Furthermore, we write $E(u) = E(t)$ for any
division $u \in \kappa^{-1}(t)$.

In section~\ref{MinimalClasses} we have introduced the minimum
$\min(n,m)$ of the tree function on the subset of complete trees $\B
\in \C(X)$ which has $m$ elements in its minimal partition
$z_{\tmin}(X)$, where the base set is $X$ with $n = \# X$. This subset
corresponds to the set $T(n,m)$ defined in section~\ref{Divisions},
and hence $\min(n,m)$ is the minimum of the descended map $E$, formula
(\ref{descend1}), in $T(n,m)$. The relation of the quantity
$\min(n,m)$ to the preoptimized minimum $p \min(n,m)$ introduced above
is as follows:
\begin{proposition}[Preoptimized minima] \label{PreoptimizedMinima}
Let $n = \# X$ and $1 \le m \le n$, then
\begin{equation}
\label{EqPreOptMin}
 p\min(n,m) \ge \min(n,m) \quad.
\end{equation}
\end{proposition}
\begin{proof}
The set of all preoptimised trees $p\O(n,m)$ in general is a proper
subset of $\M(n,m) = \left( t \circ z_{\tmin} \right)^{-1}\big( T(n,m)
\big)$. Hence, the minimum $p \min(n,m)$ of $E$ taken on $p\O(n,m)$
need not be the global minimum $\min(n,m)$ on $T(n,m)$.
\end{proof}


In the next section we shall compare the values $E(t)$ with
$E(\bar{t})$ at the optimal division $\bar{t} \in T(n,m)$.

\section{Minimality of the optimal division}
\label{MinimalityOptimalDivision}

In eq.~(\ref{EqOptDiv1}) in section~\ref{OptimalDivision} we have
defined the optimal division $\bar{t}$ of $n$ by $m$ terms. In this
section we show that the preoptimized trees for which the minimal
partition $z_{\tmin}(X)$ is optimal, or equivalently, for which $t
\circ z_{\tmin}(X) = \bar{t}$, are actually the minimal ones in
$\M(n,m)$, i.e. they lie in $\Min(n,m)$. First we show that they are
the minimal ones in the set of all preoptimized trees $p\O(n,m)$:
\begin{theorem}[Minimality of optimal partition 1]
\label{MinimalityOptimalPartition1}
Let $\bar{t} = \bar{t}(n,m)$ be the occupation number of the optimal
division of $n$ by $m$. Then
\begin{equation}
\label{EqMinOptPart1F1}
 E(t) \ge E(\bar{t}) \quad
\end{equation}
for all $t \in p\O(n,m)$.
\end{theorem}
As a consequence we have
\begin{equation}
\label{EqMinOptPart1F2}
 p \min(n,m) = E(\bar{t}) \quad,
\end{equation}
and the inverse image of $\bar{t}$ in $p\O(X)$ must therefore lie in
$\Min_p(n,m)$,
\begin{equation}
\label{EqMinOptPart2F3}
 (t\circ z_{\tmin})^{-1}(\bar{t}) \cap p\O(X) \subset \Min_p(n,m)
 \quad.
\end{equation}
\begin{proof}
Let $t = (t_1, \ldots, t_n) \in T(n,m)$. Let $u, \bar{n}$ be arbitrary
representatives of $\kappa^{-1}(t)$, $\kappa^{-1}(\bar{t})$,
respectively; this means that $u$ and $\bar{n}$ are divisions of $n$
by $m$, $u= (n_1, \ldots, n_m)$ and $\bar{n} = (\bar{n}_1, \ldots,
\bar{n}_m)$, such that eq.~(\ref{EqPreOptTrees2}) can be expressed as
\begin{equation}
\label{MinOptPart1Proof1}
 E(t) = n(m-1) + \sum_{j=1}^m G(n_j) \quad,
\end{equation}
with a similar expression for $E(\bar{t})$. It follows that
\begin{equation}
\label{MinOptPart1Proof2}
 E(t) - E(\bar{t}) = \sum_{j=1}^m \Big\{\, G(n_j) - G(\bar{n}_j) \,
 \Big\} \quad.
\end{equation}
Since $\bar{n}$ is optimal, eq.~(\ref{EqAmNonOptDiv}) in
theorem~\ref{AmountNonOptimalDivisions} immediately implies that the
RHS must be $\ge 0$, hence (\ref{EqMinOptPart1F1}) holds.
\end{proof}

\paragraph{Remark:}
The inclusion in eq.~(\ref{EqMinOptPart2F3}) is proper in
general. This means that there exist elements in $\Min_p(n,m)$ whose
associated minimal partition is {\bf not} optimal. As an example,
consider $n=6$, $X = \{1, \ldots, 6\}$, with optimal amount
\begin{equation}
\label{EqRemarkMinOptDiv1}
 G(6) = 6 + G(3) + G(3) = 6 + 5 + 5 = 16 \quad.
\end{equation}
Now compare this with the complete tree $\B = \B_{po}(2) +
\B_{po}(4)$, which is a sum of the preoptimized trees $\B_{po}(2)$ and
$\B_{po}(4)$, respectively. $\B$ is non-optimal, since the minimal
partition $z_{\tmin}(X)$ is based on the non-optimal division $(2,4)$
of $6$. The fact that $\B_{po}(4)$ is preoptimized implies that
elements $b \in z_{\min}(4)$ have cardinality $\# b =2$, and hence the
fact that the whole tree is complete implies that the subtrees
$\B(b,\B_{po}(4))$ are optimal. Thus, $G_{\B_{po}(4)} = G(4) = 8$,
whereas $G_{\B_{po}(2)} = G(2) = 2$, and thus the tree $\B$ has a
total amount of
\begin{equation}
\label{EqRemarkMinOptDiv2}
 G_{\B} = 6 + 8 + 2 = 16 \quad,
\end{equation}
which coincides with $G(6)$ in eq.~(\ref{EqRemarkMinOptDiv1}) even
though the tree $\B$ is not optimal.


The next theorem explains how $p\min(n,m)$ changes for fixed $n$ as
$m$ increases:
\begin{theorem}[Monotonicity of the preoptimized minimum]
\label{MonotonPreOptMinimum}
Let $n \ge 2$ and $1 \le m \le n$. Then
\begin{equation}
\label{MonIncPreOptMin}
 p\min(n,m+1) > p\min(n,m) \quad.
\end{equation}
\end{theorem}
\begin{proof}
The case $m=1$ yields $p\min(n,1) = 0$, whereas $p\min(n,2) = G(n) >
0$ whenever $n \ge 2$. Thus we certainly have $p\min(n,1) <
p\min(n,2)$. Therefore assume now that $m\ge 2$. Let $n$ be optimally
divided by $(m+1)$ according to $n = \nu \cdot (m+1) + r$, where $\nu
= \left[ \smallf{n}{m+1} \right]$ and $r < m+1$. The naturally ordered
representative of $\kappa^{-1}(\bar{t})$ is
\begin{equation}
\label{MonIncPreOptMinProof1}
 \bar{n} = \left(\, \underset{m+1-r}{\underbrace{\nu, \ldots, \nu}}\,
 , \, \underset{r}{\underbrace{\nu+1, \ldots, \nu+1}}\, \right)
 \quad.
\end{equation}
According to this decomposition we have from
eq.~(\ref{EqMinOptPart1F2}) in
theorem~\ref{MinimalityOptimalPartition1} and
eq.~(\ref{EqPreOptTrees2}) that
\begin{equation}
\label{MonIncPreOptMinProof2}
 E(\bar{n}) = p\min(n,m+1) = n m + (m+1-r) \cdot G(\nu) + r \cdot
 G(\nu+1) \quad.
\end{equation}
Now define a new division $u$ of $n$ into $m$ terms by
\begin{equation}
\label{MonIncPreOptMinProof3}
 u \equiv (u_1, \ldots, u_m) \equiv \left( \bar{n}_2, \ldots,
 \bar{n}_m, \bar{n}_1 + \bar{n}_{m+1} \right) \quad.
\end{equation}
The value of $E$ on any preoptimized tree whose minimal partition
$z_{\tmin}(X)$ corresponds to $u$ can be computed using
eq.~(\ref{EqPreOptTrees1}),
\begin{subequations}
\label{MonIncPreOptMinProof4}
\begin{align}
 E(u) & = n (m-1) + (m-r) \cdot G(\nu) + (r-1) \cdot G(\nu+1) +
 G(2\nu+1) \quad, \nonumber \\
 & r > 0 \quad, \label{MIPOMP4a} \\
 E(u) & = n (m-1) + (m-1) \cdot G(\nu) + G(2\nu) \quad, \nonumber \\
 & r = 0 \quad. \label{MIPOMP4b}
\end{align}
\end{subequations}
Assume first that $r>0$: In this case we have $u_m = \bar{n}_1 +
\bar{n}_{m+1} = 2\nu+1$. Use eqs.~(\ref{EqMinOptPart1F2},
\ref{MonIncPreOptMinProof2}, \ref{MIPOMP4a}) to compute
\begin{equation}
\label{MonIncPreOptMinProof6}
 p\min(n,m+1) - E(u) = n + G(\nu) + G(\nu+1) - G(2\nu+1) \quad.
\end{equation}
However, the amount $G(2\nu+1)$ in the optimal tree $\B_o(2\nu+1)$
over a set with $(2\nu+1)$ elements can be expressed using
eq.~(\ref{EqSplittingLemmaCorollary2}) in
corollary~\ref{SplittingLemmaCorollary} as
\begin{equation}
\label{MonIncPreOptMinProof7}
 G(2\nu+1) = (2\nu+1) + G(\nu) + G(\nu+1) \quad,
\end{equation}
so that (\ref{MonIncPreOptMinProof6}) yields
\begin{equation}
\label{MonIncPreOptMinProof8}
 p\min(n,m+1) - E(u) = \nu (m-1) + (r-1) \quad.
\end{equation}
Since $r \ge 1$ and $m \ge 2$, the RHS is $> 0$. -- For the case $r=0$
we use eq.~(\ref{MIPOMP4b}) to obtain in the same way as above
\begin{equation}
\label{MonIncPreOptMinProof9}
 p\min(n,m+1) - E(u) = n + 2\, G(\nu) - G(2\nu) \quad.
\end{equation}
However, $G(2\nu) = 2\nu + 2\, G(\nu)$, so that
(\ref{MonIncPreOptMinProof9}) becomes
\begin{equation}
\label{MonIncPreOptMinProof10}
 p\min(n,m+1) - E(u) = \nu (m-1) \quad,
\end{equation}
which is again greater than zero. To finish our argument we use
eqs.~(\ref{EqMinOptPart1F1}, \ref{EqMinOptPart1F2}) in
theorem~\ref{MinimalityOptimalPartition1}, which imply that
\begin{equation}
\label{MonIncPreOptMinProof11}
 E(u) \ge p\min(n,m) \quad.
\end{equation}
Now eqs.~(\ref{MonIncPreOptMinProof9} -- \ref{MonIncPreOptMinProof11})
imply the result in eq.~(\ref{MonIncPreOptMin}).
\end{proof}

The chain of inequalities in eq.~(\ref{MonIncPreOptMin}) points out
that the minimum $p\min(n,2)$ with respect to a binary optimal
division of the set $X$ is the lowest in the set of all minima
$p\min(n,m)$. It follows from eq.~(\ref{DefPreOptStar}) that
$p\min(n,2)$ is therefore the global minimum in $p\O^*(n)$. Hence, on
using the notation (\ref{DefPreOptMin1}),
\begin{corollary}[Minimality of bidivisions]
\label{MinimalityBidivisions}
$p\min(n,2)$ is the global minimum of $E$ on $p\O^*(n)$,
\begin{equation}
\label{EqMinBidiv}
 p\min(n,2) = p\min(n) \quad.
\end{equation}
\end{corollary}


The next theorem explains the role of the optimal trees in the present
context:
\begin{theorem}[Minimality of optimal division 2]
\label{MinimalityOptimalPartition2}
Let $\# X = n \ge 2$. Then the optimal trees minimize the tree
function on the set of all preoptimized trees with two elements in
$z_{\tmin}(X)$, and hence on all preoptimized trees. In symbols,
\begin{subequations}
\label{EqMinOptPart2}
\begin{equation}
 \O(X) \subset \Min_p(X,2) \subset \Min_p(X) \quad, \label{EMOP2a}
\end{equation}
and
\begin{equation}
\label{EMOP2b}
 G(n) = p\min(n,2) = p\min(n) \quad.
\end{equation}
\end{subequations}
\end{theorem}
\begin{proof}
Let $\B_o \in \O(X)$, then in particular, $\B_o$ is preoptimized, and
furthermore, the minimal partition $z_{\tmin}(X)$ is optimal, i.e.,
$(t \circ z_{\tmin})(\B_o) = \bar{t}(n,2)$. Then
(\ref{EqMinOptPart1F1}) says that $E(\bar{t}) = G(n)$ is the minimum
in $\Min_p(X,2)$. As a consequence we must have the first inclusion in
(\ref{EMOP2a}), and the first equality in (\ref{EMOP2b}) must
hold. The second inclusion in (\ref{EMOP2a}) is a consequence of
corollary~\ref{MinimalityBidivisions}.
\end{proof}


Now we come to the main theorem of this work:
\begin{theorem}[Optimal trees are globally minimal]
\label{OptimalTreesGlobalMini}
Let $\# X = n~\ge~2$. Then the optimal trees over $X$ belong to the
globally minimal trees over $X$, i.e.,
\begin{subequations}
\label{EqOptTreeGlobMin}
\begin{equation}
 \O(X) \subset \Min(X) \quad, \label{EOTGMa}
\end{equation}
and
\begin{equation}
 G(X) = \min(X) = \min(n) \quad. \label{EOTGMb}
\end{equation}
\end{subequations}
\end{theorem}
\begin{proof}
Since all trees involved in the present discussion are complete, the
tree function $E$ always coincides with the total amount $G$ of the
tree, as follows from theorem~\ref{TreeFunctionAndTotalAmount}. We
prove (\ref{EqOptTreeGlobMin}) by induction with respect to $n = \#
X$.

\underline{$n=2$:} In this case there is only one complete tree, and
hence $\O(X) = \Min(X)$ trivially.

\underline{Induction step:} We assume that
\begin{equation}
\label{EqOTGMProof1}
 \O(n') \subset \Min(n')
\end{equation}
for all $2 \le n' \le n-1$. We prove (\ref{EqOptTreeGlobMin}) for $\#X
= n$ by showing that $G_{\B} \ge G(n)$ for every complete tree $\B \in
\C(X)$ over $X$. Let $\B \in \C(X)$, let $z_{\tmin}(X) = (b_1, \ldots,
b_m)$ be the minimal partition of $X$ in $\B$. Let $u = (u_1, \ldots,
u_m) = (\# b_1, \ldots, \# b_m)$. Now apply
eq.~(\ref{EqSplittingLemmaCorollary2}) in
corollary~\ref{SplittingLemmaCorollary}:
\begin{equation}
\label{EqOTGMProof2}
 G_{\B} = n(m-1) + \sum_{j=1}^m G_{\B(b_j,X)} \quad.
\end{equation}
The subtrees need not be optimal, hence assumption
(\ref{EqOTGMProof1}) implies that
\begin{equation}
\label{EqOTGMProof3}
 G_{\B(b_j,X)} \ge G(u_j) \quad \text{for all $j=1, \ldots, m$} \quad.
\end{equation}
Thus, (\ref{EqOTGMProof2}) implies that
\begin{equation}
\label{EqOTGMProof4}
 G_{\B} \ge n(m-1) + \sum_{j=1}^m G(u_j) \equiv G_{\B'} \quad,
\end{equation}
where the RHS of the last formula defines the total amount of the {\bf
preoptimized} tree
\begin{equation}
\label{EqOTGMProof5}
 \B' \equiv \sum_{b \in z_{\tmin}(X)} \O(b) \in p\O(X,m) \quad.
\end{equation}
Now $G_{\B'} = E(t)$, where $t$ is the occupation number of $u$, $t =
\kappa(u)$; hence, by eqs.~(\ref{EqMinOptPart1F1},
\ref{EqMinOptPart1F2}) in theorem~\ref{MinimalityOptimalPartition1},
we must have $G_{\B'} \ge E(\bar{t}) = p\min(n,m)$, where $\bar{t}$ is
now the optimal division of $n$ by $m$. From (\ref{MonIncPreOptMin})
in theorem~\ref{MonotonPreOptMinimum} we know that $p\min(n,m) \ge
p\min(n,2)$. Using (\ref{EMOP2b}) in theorem
\ref{MinimalityOptimalPartition2} we have $p\min(n,2) = p\min(n) =
G(n)$. Thus, putting all these inequalities together,
\begin{equation}
\label{EqOTGMProof6}
 G_{\B} \ge G(n) = G(X) \quad.
\end{equation}
which proves the theorem.
\end{proof}

\section{Mean path amount and quadratic deviation}
\label{MeanPathQuadDeviation}

From section~\ref{OptimalAmount}, formula~(\ref{EqOptAmount}), we
immediately see that the mean path amount $\smallf{1}{n} \sum e_i$
will be close to $\lg(n)$. We can make this statement more precise:
\begin{definition}[Mean path amount] \label{DefMeanPathAmount}
The {\em mean path amount} $\b{e}_{\B}$ in the tree $\B(X)$ is defined
to be
\begin{equation}
\label{EqDefMeanPathAm}
 \b{e}_{\B} \equiv \min \big\{\, \eta \in \NN \, \big|\, \eta \cdot n
 \ge G_{\B} \,\big\} \quad.
\end{equation}
\end{definition}
Thus
\begin{equation}
\label{EqMeanPathAm1}
 G_{\B} = \b{e}_{\B} \cdot n -r \quad, \quad \text{with $0 \le r< n$}
 \quad.
\end{equation}
In particular:
\begin{proposition}[Mean path amount in optimal trees 1]
\label{MeanPathAmountOptimalTrees1}
In an optimal tree $\B = \B_o$,
\begin{equation}
\label{EqMeanPathAm2}
 \b{e}_{\B_o} = \left\{ \begin{array}{lcl} \lg(n) & , & n \in \BB \\
 \lg(n) + 1 & , & n \not\in \BB \end{array} \right. \quad.
\end{equation}
\end{proposition}
\begin{proof}
If $n \in \BB$ then $\b{e}_{\O} = \lg(n)$ follows immediately from
(\ref{EqOptAmount}), since $n = 2^{\lg(n)}$ in this case. 

Now assume that $n \not\in \BB$. From the maximum property of $\lg(n)$
it follows that $n < 2^{\lg(n)+1}$. This can be rearranged to give
\begin{equation}
\label{EqMPA1Proof1}
 2 \left[ n - 2^{\lg(n)} \right] < n \quad.
\end{equation}
If we add $n \lg(n)$ on both sides of this inequality we obtain
\begin{equation}
\label{EqMPA1Proof2}
 n \cdot \lg(n) + 2 \big[ n - 2^{\lg(n)} \big] < n \left[ \lg(n)
 + 1 \right] \quad.
\end{equation}
However, the LHS is just $E_{\B_o}$ according to
formula~(\ref{EqOptAmount}), and hence
\begin{equation}
\label{EqMPA1Proof3}
 E_{\B_o} < n \big[ \lg(n) + 1 \big] \quad.
\end{equation}
On the other hand, the same formula~(\ref{EqOptAmount}) says that
\begin{equation}
\label{EqMPA1Proof4}
 E_{\B_o} \ge n \cdot \lg(n) \quad.
\end{equation}
Since $n \not\in \BB$ we have $n = 2^{\lg(n)} + r$, where $0 < r <
2^{\lg(n)}$. In this case the inequality in
formula~(\ref{EqMPA1Proof4}) becomes proper, and thus $\lg(n)+1$
satisfies the minimum property in formula (\ref{EqDefMeanPathAm}),
definition~\ref{DefMeanPathAmount}.
\end{proof}


We now come back to eq.~(\ref{EqMeanPathAm1}), $G_{\B} = ( \b{e}_{\B}
\cdot n -r)$, where $r < n$. Now let us define the $n$-tupel
\begin{equation}
\label{EqMeanPathAm3}
 (\b{e}_1, \ldots, \b{e}_n) \equiv \Big(\, \underset{n-r}{
 \underbrace{ \b{e}_{\B}, \ldots, \b{e}_{\B}}}\, , \, \underset{r}{
 \underbrace{ \b{e}_{\B}-1, \ldots, \b{e}_{\B}-1 }} \, \Big) \quad.
\end{equation}
Thus, for any tree $\B$ over $X$ (which need not be optimal) we have
\begin{equation}
\label{EqMeanPathAm4}
 G_{\B} = \sum_{i=1}^n e_i = \sum_{i=1}^n \b{e}_i = (n-r) \cdot
 \b{e}_{\B} + r \cdot ( \b{e}_{\B} -1 ) \quad,
\end{equation}
where we have used eqs.~(\ref{EqMeanPathAm1}, \ref{EqMeanPathAm3}).
Introducing the $n$-tupel of deviations
\begin{equation}
\label{EqMeanPathAm5}
 \Delta e \equiv ( \Delta e_1, \ldots, \Delta e_n ) \equiv ( e_1 -
 \b{e}_1, \ldots, e_n - \b{e}_n ) \quad,
\end{equation}
and the {\it total quadratic deviation in $\B(X)$} by
\begin{equation}
\label{EqMeanPathAm6}
 \sigma^2_{\tot}(\B) \equiv \sum_{i=1}^n \left( \Delta e_i \right)^2
 \quad,
\end{equation}
we find on using (\ref{EqMeanPathAm4}) that
\begin{equation}
\label{EqMeanPathAm7}
 \sigma^2_{\tot} = \sum_{i=1}^n \left( e_i^2 - \b{e}_i^2 \right) + 2
 \sum_{i=n-r+1}^n \Delta e_i \quad.
\end{equation}

We now present some statements about the mean path amount in optimal
trees. In every tree $\B$, the elements $b_1, \ldots, b_{\kappa}$ in
the maximal partition $z_{\tmax}(X)$ can be labelled so that the
associated path amounts are monotonically decreasing, $e_1 \ge e_2 \ge
\cdots e_{\kappa}$. In particular, if $\bar{e}_{\B_o}$ is the mean
path amount in the optimal tree $\B_o$ as defined in
eq.~(\ref{EqDefMeanPathAm}), then we have:
\begin{theorem}[Mean path amount in optimal trees 2]
\label{MeanPathAmountOptimalTrees2}
Let $G(n)$ be the amount of the optimal tree $\B_o$ with $n = \#
X$. Let $r \equiv n \cdot \bar{e}_{\B_o} - G(n)$. Then, if $n \in
\BB$,
\begin{subequations}
\label{EqMeanPathAm8}
\begin{equation}
\label{EMPAforma}
 e_i = \lg(n)
\end{equation}
for all $i=1, \ldots, n$, whereas for $n \not\in \BB$,
\begin{equation}
\label{EMPAformb}
 e_i = \left\{ \begin{array}{lcl} \lg(n)+1 & , & i=1, \ldots, n-r \\
 \lg(n) & , & i= n-r+1 , \ldots, n \end{array} \right. \quad.
\end{equation}
\end{subequations}
Hence, in any case,
\begin{equation}
\label{EMPAformc}
 e_i = \bar{e}_i \quad, \quad i=1, \ldots, n \quad,
\end{equation}
where the tupel $(\bar{e}_1, \ldots, \bar{e}_n)$ was defined in
eq.~(\ref{EqMeanPathAm3}).
\end{theorem}
\begin{proof}
We first prove (\ref{EMPAforma}) by induction with respect to
$\lg(n)$: For $n=1, 2$, corresponding to $\lg(n)= 0, 1$,
(\ref{EMPAforma}) is trivially satisfied. Now choose $k \equiv \lg(n)
> 1$ and suppose that (\ref{EMPAforma}) is true for $k-1$. From
formula~(\ref{DefAmountOfB}) we know that the amount $e(b)$ of a path
in any optimal tree is equal to $\# q(b) -1$, since $m(b)= 2$ for all
non-terminal elements, while $m(b)=1$ for all terminal ones. Thus, the
amount $e(b)$ of any terminal element is $1$ plus the amount of the
same element in the subtree $\B(a,X)$, where $a \in z_{\tmin}(X)$ and
$b \subset a$. By assumption, $n= 2^k$, hence the minimal partition of
$X$ contains two elements $a_1$ and $a_2$ both of which must have the
same cardinality $\# a_1 = \# a_2 = 2^{k-1}$. Let $b$ be any terminal
element in the tree such that $b \subset a_1$, say. Then
$e_{\B(a,X)}(b) = k-1$ by assumption. In the full tree, the path
length of the same element $b$ is greater by just one, hence
$e_{\B_o}(b) = k = \lg(n)$, which confirms (\ref{EMPAforma}).

Now assume $n \not\in \BB$. We first show: The path lengths $e_i$ can
mutually differ at most by $\pm 1$,
\begin{equation}
\label{EMPAproof1}
 \left| e_i - e_j \right| \in \{0, 1\} \quad.
\end{equation}
We prove this statement by induction with respect to $n$: For $n=1$
and $n=2$ the path lengths in the optimal trees are $0$ and $1$,
respectively, and hence (\ref{EMPAproof1}) is satisfied. For $n=3$ the
path amounts in the optimal tree are $e_1 = e_2 = 2$ and $e_3 =1$;
again, (\ref{EMPAproof1}) is satisfied. Now let $n \ge 4$ and assume
that statement (\ref{EMPAproof1}) is true for all $1 \le n' \le
n-1$. Let $b, b'$ be any two elements in the maximal partition
$z_{\tmax}$ of $X$. Let $a, a'$ denote those elements in the minimal
partition $z_{\tmin}(X)$ for which $b \subset a$ and $b' \subset a'$
(this includes the possibility that $a=a'$). Then $\# a, \# a' < n$,
and the induction assumption applies to the path amounts in the
optimal subtrees $\B(a,X)$ and $\B(a',X)$: Namely, since $e(b) =
e_{\B(a,X)}(b) + 1$ and $e(b') = e_{\B(a',X)}(b') + 1$ we must have
\begin{equation}
\label{EMPAproof2}
 \Big|\, e(b) - e(b')\, \Big| = \Big|\, e_{\B(a,X)}(b) -
 e_{\B(a,X)}(b') \, \Big| \in \{0, 1\} \quad.
\end{equation}
This proves formula~(\ref{EMPAproof1}).

From formula~(\ref{EMPAproof1}) we infer that there exist integers
$\alpha$ and $k$ such that
\begin{equation}
\label{EMPAproof2zw3}
 \alpha (k+1) + (n-\alpha) k = G(n) \quad,
\end{equation}
where $0 \le \alpha < n$. If $\alpha$ were zero we would have $G(n)= n
k$, and together with eq.~(\ref{EqOptAmount}) it would follow that
\begin{equation}
\label{EMPAproof3}
 n \Big[ \lg(n) + 2-k\Big] = 2^{\lg(n)+1} \quad.
\end{equation}
By means of prime number factorisation of the factors on the left-hand
side we conclude that $n$ must take the form $n =2^K$ for some integer
$K$, thus implying $n \in \BB$, which contradicts the initial
assumption. Hence we really have $\alpha> 0$. Now
eq.~(\ref{EMPAproof2zw3}) can be written in the form
\begin{equation}
\label{EMPAproof4}
 G(n) = n\cdot k + \alpha \quad, \quad 0 < \alpha < n \quad.
\end{equation}
If $n$ and $G(n)$ are given numbers we can consider (\ref{EMPAproof4})
as an equation for the unknowns $\alpha, k$. If the restriction $0 <
\alpha < n$ is upheld, then the solution for the pair $(\alpha, k)$ is
unique. Now consider formula (\ref{EqOptAmount}) for $G(n)$,
\begin{equation}
\tag{\ref{EqOptAmount}}
 G(n) = n \cdot \lg(n) + 2 R \quad, \quad R = n - 2^{\lg(n)} \quad.
\end{equation}
From formula~(\ref{ESDb}) in lemma~\ref{StandardDecomposition} we
know that $2R < n$. Thus, the pair
\begin{equation}
\label{EMPAproof5}
 \alpha = 2R \quad, \quad k = \lg(n) \quad
\end{equation}
is the unique solution to the system (\ref{EMPAproof4}). As a
consequence, amongst the $e_i$ there must be $(n-2R)$ occurrences of
$\lg(n)$ and $2R$ occurrences of $\lg(n)+1$,
\begin{equation}
\label{EMPAproof6}
 e_i = \left\{ \begin{array}{lcl} \lg(n) + 1 & , & i = 1, \ldots, 2R
 \\ \lg(n) & , & i = 2R + 1 , \ldots, n \end{array} \right. \quad.
\end{equation}
It remains to show that $n-r = 2R$: To this end we write down
formula~(\ref{EqMeanPathAm1}) for the case at hand, i.e., an optimal
tree with $n \not\in \BB$, in which case (\ref{EqMeanPathAm2})
applies,
\begin{equation}
\label{EMPAproof7}
 G(n) = \big[\, \lg(n)+1\, \big] \cdot n - r \quad.
\end{equation}
Comparison of (\ref{EMPAproof7}) with formula~(\ref{EqOptAmount})
gives $2R = n-r$, hence (\ref{EMPAformb}) is proved.
\end{proof}

\section{Isomorphic trees} \label{IsomorphicTreeStructures}

In this section we formulate a notion of structural similarity between
trees $\B$ and $\B'$ which no longer need to be defined over the same
set $X$. This will lead to an appropriate notion of isomorphism of
trees.

Consider a tree $\B$ over a set $X$. The structure of the tree $\B$ is
captured in the set of its nodes $b$, and the degree of splitting
$m(b)$ associated with each node. The particular nature of the
underlying set $X$, just as the particular value $n(b)$ of the
cardinality of the nodes, is not a primary structure-determining
element. To see this we can construct different trees from the given
tree $\B$ which exhibit the same structure: To this end consider the
maximal partition $z_{\tmax}(X,\B) = \{c_1, \ldots, c_K\}$ of $X$ in
$\B$. The elements $c_i \in z_{\tmax}(X,\B)$ are terminal in this tree
and are never partitioned further; this means that their "internal
structure" is immaterial, as far as the tree $\B$, and its internal
structure, are concerned. Now consider any collection $\{c'_1, \ldots,
c'_K\}$ of non-empty, mutually disjoint sets $c'_i$ with $i= 1,
\ldots, K$, and let $X' \equiv \bigcup_i c'_i$. We can think of
constructing a new tree $\B'$ by replacing every terminal element
$c_i$ in the old tree $\B'$ by the corresponding element $c'_i$. Then
there is a 1--1 relation between nodes $b \in \B$ and $b' \in \B'$;
moreover, the minimal partitions $z_{\tmin}(b',\B'(X'))$ and
$z_{\tmin}(b,\B(X))$ are the same for all nodes $b$ and $b'$ which
correspond to each other. In particular, the degree of splitting
$m(b')$ is the same as $m(b)$ for such nodes. Obviously, both trees
have the same cardinality, $\# \B' = \# \B$. This idea can be made
precise in the following
\begin{definition}[Isomorphism of trees] \label{IsomorphismTrees}
Let $\B$ and $\B'$ be trees over sets $X$ and $X'$ with the same
cardinality, $\# B = \# \B'$. $\B$ and $\B'$ are {\it isomorphic} if
there exists a bijection $i: \B \rightarrow \B'$ such that
\begin{equation}
\label{EqIsoTrees}
 m \circ i (b) = m(b) \quad \text{for all $b \in \B$} \quad.
\end{equation}
\end{definition}
For a given pair $\B$ and $\B'$ of trees there can exist more than one
isomorphism.
\begin{proposition}[Paths in isomorphic trees]
\label{PathAmountInIsomorphicTrees}
Let $i: \B \rightarrow \B'$ be an isomorphism of trees. Then the path
assignment $b \mapsto q(b)$ commutes with $i$,
\begin{subequations}
\begin{equation}
\label{EqPathAmountIso1}
 q \circ i = i \circ q \quad.
\end{equation}
As a consequence, isomorphic trees have the same path amounts,
\begin{equation}
\label{EqPathAmountIso2}
 e_{i\B}(i(b)) = e_{\B}(b) \quad \text{for all $b \in \B$} \quad.
\end{equation}
\end{subequations}
\end{proposition}
\begin{proof}
Since isomorphic trees have the same basic structure in the sense that
they have the same number of nodes, and each node has the same degree
of splitting, the paths in isomorphic trees have the property that
\begin{equation}
\label{EqPAIproof1}
 q(i(b)) = i(q(b)) \quad \text{for all $b \in \B$} \quad,
\end{equation}
which gives (\ref{EqPathAmountIso1}). Furthermore, $m \circ i = i
\circ m$ by definition of isomorphism. Then the statement
(\ref{EqPathAmountIso2}) follows immediately from
eq.~(\ref{DefAmountOfB}) in definition~\ref{AmountOfNode}.
\end{proof}

For a given tree $\B$ we can think of the category $[\B]$ of
isomorphic trees. Even if $\B$ is defined over a finite set $X$,
general elements of $[\B]$ need no longer share this property; all
that is required is that they have the same number of nodes and the
same degree of splitting $m(b)$ as the original tree $\B$ at each node
$b$. From proposition~\ref{PathAmountInIsomorphicTrees} we learn that
all trees in the category $[\B]$ have the same path lengths $o(b) = \#
q(b)$, and the same path amounts $e(b)$. In general they differ in the
total amount $E_{\B}$, however.

There exists a stronger form of isomorphism which can be defined for
trees which are built over the same base set $X$:
\begin{definition}[Equivalent trees] \label{EquivalentTrees}
Let $\B$ and $\B'$ be trees over the same base set $X$. $\B$ and $\B'$
are said to be {\em equivalent} if they are isomorphic and share the
same maximal partition of $X$,
\begin{equation}
\label{EqMaxPartIsoTree1}
 z_{\tmax}(X,\B) = z_{\tmax}(X,\B') \quad.
\end{equation}
\end{definition}
Let us assume that $X$ is finite. Then there exists an integer $K$
such that $z_{\tmax}(X,\B) = \{c_1, \ldots, c_K\}$ and
$z_{\tmax}(X,\B') = \{ c'_1, \ldots, c'_K \}$. Equivalence of $\B$ and
$\B'$ then means that there exists a permutation $\pi$ of $K$ elements
such that
\label{EqMaxPartIsoTree2}
\begin{equation}
\label{EMPIT2a}
 c'_j = c_{\pi(j)} \quad \text{for all $j=1, \ldots, K$} \quad.
\end{equation}
But $\B$ and $\B'$ are isomorphic, hence $c'_j = i(c_j)$, where $i: \B
\rightarrow \B'$ is an appropriate isomorphism. By
eq.~(\ref{EqPathAmountIso2}), the path amounts are related by
$e_{\B'}(c'_j) = e_{\B}(c_j) \equiv e_j$. Let $w_j \equiv n(c_j)$;
then it follows from eq.~(\ref{EqTreeFunctionAndTotalAmount1}) in
theorem~\ref{TreeFunctionAndTotalAmount} that the following statements
are true:
\begin{theorem}[Tree function on equivalent trees]
\label{TreeFunctionEquivalentTrees}
Let $\B$ and $\B'$ be equivalent trees over the finite set $X$. Then
\begin{equation}
\label{EqTFET1}
\left.
\begin{aligned}
 E_{\B} & = \sum\limits_{j=1}^K w_j \cdot e_j \quad, \\
 E_{\B'} & = \sum\limits_{j=1}^K w_{\pi(j)} \cdot e_j \quad,
\end{aligned} \qquad
\right\}
\end{equation}
where $\pi$ is a permutation of $K$ objects.
\end{theorem}

Each category $[\B]$ contains preferred elements $\S$ which we can
construct as follows: Let $\B \in [\B]$, and consider the maximal
partition $z_{\tmax}(X,\B)= \{c_1, \ldots, c_K\}$ as before. Now
define the set $X_S \equiv \{1, \ldots, K\}$. $\S$ is now defined to
be a tree over $X_S$, isomorphic to $\B$, and is obtained by replacing
every terminal element $c_i$ in the maximal partition
$z_{\tmax}(X,\B)$ by the one-element set $\{i\}$. More generally,
every node $b = \bigcup_{i_1, \ldots, i_{\kappa}} c_{i_{\kappa}}$ is
replaced by the set $\{i_1, \ldots, i_{\kappa}\}$. By construction,
the tree so obtained has the same number of nodes as $\B$ and has the
same degree of splitting at each node. However, everything about the
internal structure of the terminal elements $c_i$ in the maximal
partition of $\B$ has been stripped away, so that the tree now
incorporates nothing more than the inherent structure which is shared
by all trees in the category $[\B]$. It is then befitting to call such
a tree $\S$ a "skeleton" of $\B$, and hence a skeleton in the
respective category. The defining criterion of a skeleton is the fact
that all terminal elements are one-element sets, i.e., that $\S$ is
complete:
\begin{definition}[Skeleton] \label{Skeleton}
A tree $\S \in [\B]$ which is complete is called a {\it skeleton} in
the category $[\B]$.
\end{definition}
Thus, it is the skeletons which embody the inherent structure in the
category $[\B]$.
\begin{proposition}[Tree function on isomorphic trees]
\label{TreeFunctionIsomorphicTrees}
Let $\B$ be a tree over $X$. Let $\S$ be a skeleton in the category
$[\B]$. Then
\begin{equation}
\label{EqTreeFunctionIsomorphicTrees}
 E_{\B} = \sum_{c \in z_{\tmax}(X,\B)} n(c) \cdot e_{\S}\big( i(c)
 \big) \quad,
\end{equation}
where $i: \B \rightarrow \S$ is the associated isomorphism.
\end{proposition}
\begin{proof}
From eq.~(\ref{EqPathAmountIso2}) in
proposition~\ref{PathAmountInIsomorphicTrees} we know that $e_{\B}(c)
= e_{\S}\big( i(c) \big)$ for all $c \in z_{\tmax}(X,\B)$; if this is
inserted into eq.~(\ref{EqTreeFunctionAndTotalAmount1}) in
theorem~\ref{TreeFunctionAndTotalAmount},
(\ref{EqTreeFunctionIsomorphicTrees}) follows.
\end{proof}

It follows that the tree function $E$, when restricted to the category
$[\B]$, takes its minimum on the skeletons $\S \in [\B]$, since for
these, $n(c) =1$ for all $c \in z_{\tmax}(X_S,\S)$.

\section{Restricted minimal problems} \label{RestrictedMinimalProblems}

In the previous sections we have solved the problem of minimizing the
tree function $E$ on the set of all unconstrained complete trees over
the set $X$. By unconstrained we mean that no conditions on the
possible trees $\B$ over $X$ were imposed other than requiring that
$\B$ must not be trivial. We now investigate how to extend the
framework we have worked in so far in order to obtain tree functions
that contain expressions like $\sum p_i \log_2(p_i)$ in the functional
form of their minimal value, when restricted to certain classes of
tree structures over $X$.

Amongst the countless ways to constrain the set of admissible trees we
shall consider the following two cases only: For a given partition $z$
of the base set $X$ we first study the set of all trees preserving the
partition $z$; and then, the set of all trees containing $z$.

\subsection{Trees preserving a partition}
\label{TreesPreservingPartition}

A complete tree has a maximal partition of $X$ which is complete,
i.e., the elements of $z_{\tmax}(X)$ are comprised by the one-element
subsets $\{x\}$ for $x \in X$. Trivially, every partition $z$ of $X$
preserves $z_{\tmax}(X)$ in the sense that $z_{\tmax}$ is a refinement
of every partition $z$ of $X$. We now generalize this reasoning to the
case where $z_{\tmax}(X)$ is no longer complete: We want to prescribe
a partition $z$ of $X$ such that the relation $z' \preceq z$ is true
for all $z' \in \zeta(X,\B)$ compatible with $\B$. In particular, for
the maximal partition of $X$ in $\B$ we must have $z_{\tmax}(X)
\preceq z$. If such a relation is true we shall say that {\it the tree
$\B(X)$ preserves the partition $z$}. In general, the prescribed
partition $z$ that is preserved by the admissible trees $\B$ need not
be an element of $\zeta(X,\B)$ itself; in this case it induces a
non-trivial partition on at least one of the elements $b \in
z_{\tmax}(X)$ which are terminal in $\B$, so that the resulting
refinement $z$ of $z_{\tmax}(X)$ is compatible with the resulting
extension of $\B$. Alternatively, we can have $z= z_{\tmax}(X)$; in
this case, $z$ is the most refined partition compatible with the tree
$\B$. These ideas lead to
\begin{definition}[$z$-preserving, $z$-complete trees]
\label{ZPreservingZCompleteTrees}
Let $\B \in \M(X)$, let $z \in \Z(X)$ be a partition of $X$. $\B$ is
called {\em $z$-preserving} if $z' \preceq z$ for all $z' \in
\zeta(X,\B)$. $\B$ is called {\em $z$-complete} if $z_{\tmax}(X,\B) =
z$.
\end{definition}
It is clear that, without further conditions, it makes no sense to ask
for the minimum of $E$ on the set of all $z$-preserving trees over
$X$, as the answer is trivial: If $z$ is given, the minimum is taken
on the trivial tree $\B = \{X\}$, since the trivial tree preserves
every partition. And even if this trivial solution is excluded, then
the tree function $E$ takes its minimum on any binary non-trivial
partition of $X$ which preserves $z$; the associated value of the
minimum can be inferred immediately from
eq.~(\ref{EqTreeFunctionAndTotalAmount1}) to be $E = n$.

However, a meaningful minimal problem can be given on the smaller set
of $z$-complete trees, which we shall denote by $\C(z)$. The minimum
of $E$ on $\C(z)$ will be denoted by $\min\nolimits_+(z)$. The subset
of all trees in $\C(z)$ on which $E$ actually takes the minimum will
be written as $\Min_+(z)$; it coincides with the set $E^{-1}\big(
\min\nolimits_+(z) \big) \cap \C(z)$.

In the present work we shall not attempt to solve this minimal
problem; however, we provide a necessary condition which arises in the
course of its study:
\begin{proposition} \label{NecessaryCondition}
Let $z = \{c_1, \ldots, c_K\}$ be the common maximal partition in
$\C(z)$. Let $\B \in \Min_+(z)$. Then,
\begin{equation}
\label{EqNecCond1}
 w_i < w_j \quad \Rightarrow \quad e_i \ge e_j \quad,
\end{equation}
where $w_k \equiv n(c_k)$, and $e_k$ are the path amounts of $c_k$ in
$\B$.
\end{proposition}
\begin{proof}
Let $\B \in \Min_+(z)$ with path amounts $e_k$, $k =1, \ldots,
K$. From eq.~(\ref{EqTreeFunctionAndTotalAmount1}) in
theorem~\ref{TreeFunctionAndTotalAmount} we know that
\begin{equation}
\label{EqNecCondProof1}
 E_{\B} = \sum\limits_{k=1}^K w_k \cdot e_k \quad.
\end{equation}
Assume that there exist $i \neq j$ with $w_i < w_j$ such that $e_i <
e_j$. We define a new tree $\B'$ by the statements: (1) $\B'$ is
equivalent to $\B$; and (2) the maximal partition $z_{\tmax}(X,\B') =
\{ c'_1, \ldots, c'_K \}$ of $X$ in $\B'$ is such that
\begin{equation}
\label{EqNecCondProof2}
 c'_k = \tau(i,j)\, c_k \quad,
\end{equation}
where $\tau(i,j)$ is the transposition of $i$ and $j$. The path
amounts are the same by definition of equivalence,
\begin{equation}
\label{EqNecCondProof3}
 e_{\B'}(c_k) = e_{\B}(c_k) = e_k \quad,
\end{equation}
hence the tree function on $\B'$ takes the value
\begin{equation}
\label{EqNecCondProof4}
 E_{\B'} = \sum\limits_{k \neq i,j} w_k \cdot e_k + w_j \cdot e_i +
 w_i \cdot e_j \quad.
\end{equation}
It follows that
\begin{equation}
\label{EqNecCondProof5}
 E_{\B'} - E_{\B} = ( w_j -w_i) (e_i -e_j) < 0 \quad,
\end{equation}
and hence $E_{\B'} < E_{\B}$, which contradicts the minimal property
of $\B$. Thus, the initial assumption was wrong, and implication
(\ref{EqNecCond1}) must be true.
\end{proof}

\subsection{Trees containing a partition}
\label{TreesContainingPartition}

Another construction is the set of trees containing the partition $z$:
The idea is that we can constrain trees by requiring that all
admissible trees contain the elements of a prescribed partition; this
leads to the
\begin{definition}[Trees containing a partition]
\label{DefTreesContainingPartition}
Let $z \in \Z(X)$. The tree $\B$ is said to {\em contain} the
partition $z$ if
\begin{subequations}
\label{TreeContainZ1}
\begin{align}
 z & \subset \B(X) \quad \Leftrightarrow \label{TCZ1a} \\
 a & \in \B(X) \quad \text{for all $a \in z$} \label{TCZ1b}
\end{align}
\end{subequations}
is true. 
\end{definition}
Without further conditions, the minimum of $E$ will always be taken on
a tree for which the prescribed partition $z$ coincides with the
maximal partition in this tree; for, any further splitting, beyond the
nodes $a \in z$, can only increase the value of $E$. It follows that
there are two possibilities for meaningful minimal problems: (1) We
require that, for all admissible trees, the maximal partition
$z_{\tmax}(X,\B)$ agrees with $z$; or, (2) we require that all
admissible trees are {\it complete}. Clearly, case (1) agrees with the
minimal problem on the set $\C(z)$ of all $z$-complete trees as
discussed in the last paragraph after
definition~\ref{ZPreservingZCompleteTrees}; the minimum of $E$ is
$\min\nolimits_+(z)$, and the set of all trees on which the minimum is
taken is $\Min_+(z)$. Case (2) defines another meaningful minimal
problem which nevertheless can be traced back to case (1): Suppose
that $\B$ is an admissible tree with respect to case (2); then $\B$
can be regarded as the completion of a reduced tree $\B'$, where $\B'$
is an element in the set $\C(z)$ of $z$-complete trees. It is then
clear that minimal trees with respect to case (2) are those for which
the reduction $\B'$ is minimal in $\C(z)$, in other words, $\B' \in
\Min_+(z)$, and for which the subtrees $\B(b_i,X)$, $b_i \in z$, are
optimal. The minimal value of the tree function $E$ in case (2) then
will be a sum of $\min\nolimits_+(z)$ and another sum over expressions
$w_i \lg(w_i) + 2 [ w_i - 2^{\lg(w_i)} ]$, where $w_i = n(b_i)$, $b_i
\in z$,
\begin{equation}
\label{TreeContainZ2}
 E_{\tmin} = \min\nolimits_+(z) + \sum_{i=1}^K \bigg\{\, w_i \cdot
 \lg(w_i) + 2 \cdot \Big[ w_i - 2^{\lg(w_i)} \Big] \, \bigg\} \quad.
\end{equation}
Here we have again assumed that $\# z \equiv K$.

\subsection{Trees with a prescribed minimal partition $z_{\tmax}(X)$}
\label{TreesPrescribedMinimalPartition}

Another minimal problem can be obtained by prescribing the minimal
partition $z_{\tmin}(X) = z$ of $X$ and demanding that all trees in
this class be complete. We shall denote the associated class of trees
by $\C_-(z)$. The minimum of $E$ taken in $\C_-(z)$ will be written as
$\min\nolimits_-(z)$, while the subset of $\C_-(z)$ on which this
minimum is actually taken will be denoted by $\Min_-(z)$; the latter
coincides with the intersection $E^{-1}\big( \min\nolimits_-(z) \big)
\cap \C_-(z)$.

The solution to this minimal problem is readily found: Let us suppose
that the minimal partition of $X$ is prescribed to be
\begin{subequations}
\label{PreMinPart1}
\begin{align}
 z_{\tmin}(X) & = z = \big\{ c_1, \ldots, c_K \big\} \quad,
 \label{PMP1a} \\
 w_i & = n(c_i) \quad, \quad i=1, \ldots, K \quad. \label{PMP1b}
\end{align}
\end{subequations}
Since all admissible trees are complete, by
eq.~(\ref{EqTreeFunctionAndTotalAmount2}) in
theorem~\ref{TreeFunctionAndTotalAmount} the tree function $E$ on
$\M_-(z)$ coincides with the total amount function
$G$. Eq.~(\ref{EqSplittingLemmaCorollary2}) in
corollary~\ref{SplittingLemmaCorollary} then implies that
\begin{equation}
\label{PreMinPart2}
 E_{\B} = n(K-1) + \sum\limits_{i=1}^K G_{\B(c_i,X)} \quad,
\end{equation}
for all $\B \in \M_-(z)$. It follows that $E_{\B}$ becomes minimal if
all subtrees $\B(c_i,X)$ become optimal; in other words, if
\begin{equation}
\label{PreMinPart3}
 G_{\B(c_i,X)} = G(w_i) \quad \text{for all $i = 1, \ldots, K$} \quad.
\end{equation}
But the values of the quantities $G(w_i)$ are given in
eq.~(\ref{EqOptAmount}) of theorem~\ref{AmountOptimalTrees},
\begin{equation}
\label{PreMinPart4}
 G(w_i) = w_i \cdot \lg(w_i) + 2\cdot \big[ w_i - 2^{\lg(w_i)} \big]
 \quad.
\end{equation}
On inserting (\ref{PreMinPart3}, \ref{PreMinPart4}) into
(\ref{PreMinPart2}) we have proven:
\begin{theorem}[Minimal problem on $\C_-(z)$\,]
\label{MinimalProblemOnMMinusZ}
The minimum of $E$ on $\C_-(z)$ is equal to
\begin{equation}
\label{EqMinProbOnMMinusZ1}
 \min\nolimits_-(z) = n(K-1) + \sum_{i=1}^K \Bigg\{\, w_i \cdot
 \lg(w_i) + 2 \cdot \big[ w_i - 2^{\lg(w_i)} \big]\, \Bigg\} \quad.
\end{equation}
\end{theorem}
The set $\Min_-(z)$ contains those trees which are sums of optimal
trees over the quantities $w_i$,
\begin{equation}
\label{EqMinProbOnMMinusZ2}
 \B = \sum_{i=1}^K \B_{o}(c_i) \Rightarrow  \B \in \Min_-(z) \quad.
\end{equation}
We can rewrite the result (\ref{EqMinProbOnMMinusZ1}) in such a way
that probability-like quantities $\smallf{w_i}{n} \in \RR$ appear:
\begin{equation}
\label{EqMinProbOnMMinusZ3}
\begin{aligned}
 & \frac{1}{n}\, \min\nolimits_-(z) - \lg(n) = (K-1)\; + \\
 & + \sum\limits_{i=1}^K \frac{w_i}{n} \, \Big[ \lg(w_i) - \lg(n)
 \Big] + \frac{1}{n} \sum\limits_{i=1}^K 2\cdot \Big[ w_i -
 2^{\lg(w_i)} \Big] \quad.
\end{aligned}
\end{equation}
The quantity $\left[ \lg(w_i) - \lg(n) \right]$ is evidently an
approximation to $\log_2(\smallf{w_i}{n})$, so that the right-hand
side of (\ref{EqMinProbOnMMinusZ3}) contains an integer approximation
to the Shannon-Wiener entropy with respect to the "probabilities" $p_i
= \smallf{w_i}{n}$, where $i=1, \ldots, K$.

\section{Tree structures and neighbourhood topology}
\label{TreesAndNeighbourhoodTopology}

Finally, we want to put forward arguments to show how tree structures
define a neighbourhood topology on the underlying set $X$. We now
allow the set $X$ to have arbitrary cardinality; in particular, $X$
can be non-countable. We recall that the path $q(b)$ of a node $b \in
\B(X)$ was defined to be the set of all elements $b'$ in $\B$
containing $b$ as a subset. We now extend this definition so as to
speak of the path of any single element $x \in X$ in the base set: For
every $x \in X$ there exists precisely one terminal element $b_x \in
\B$ such that $x \in b$; we can then decree that the path of the
element $x$ in the tree $\B$ be the path of the associated terminal
element, and this assignment will be unique. Thus,
\begin{definition}[Path of points in base set $X$]
\label{PathPointsBaseSet}
Let $\B \in \M(X)$ be a given tree over the base set $X$. Let $x \in
X$, let $b_x$ be the uniquely determined terminal node in $\B$ which
contains $x$ as an element. Then the {\em path of $x$ in $\B$} is
defined by
\begin{equation}
\label{DefPathPoint}
 q(x) \equiv q(\{b_x\}) \quad.
\end{equation}
\end{definition}
If the degree of splitting $m(b)$ remains finite at every node $b \in
\B$, the path of $x$ will be a countable subset of the tree $\B$.
\begin{proposition}[The path of points] \label{ThePathOfPoints}
Let $\B$ be a given tree over $X$. Then
\begin{equation}
\label{EqThePathOfPoints}
 b \in q(x) \Leftrightarrow x \in b \quad.
\end{equation}
\end{proposition}
\begin{proof}
Let $x \in X$ and assume that $b \in q(x)$. There exists a unique $b'
\in z_{\tmax}(X,\B)$ such that $x \in b'$. Thus, $q(b') = q(x)$, and
therefore $b \supset b' \ni x$, which proves the implication from
left-to-right.

Conversely, let $b'$ be the unique terminal element in
$z_{\tmax}(X,\B)$ such that $x \in b'$. Then $x \in b$ implies that $b
\cap b' \neq \emptyset$. Now, axiom (A2) in section~\ref{Axioms}
implies that either $b \subsetneqq b'$ or $b' \subset b$. The first
inclusion cannot be true since $b'$ is a terminal element; thus, $b'
\subset b$, hence it follows that $b \in q(b') = q(x)$.
\end{proof}

We now show that the given tree structure $\B$ over $X$ defines a
neighbourhood topology on $X$. We recall \cite{HeuserII,Brown} that a
neighbourhood topology $\N$ assigns a collection $\N(x)$ of distinct
subsets $N$ of $X$ to every point $x$ in $X$; $\N$ is just the
collection of all $\N(x)$. The elements $N \in \N(x)$, which are
subsets of $X$, are called {\it neighbourhoods of $x$ in the topology
$\N$}, if they satisfy the axioms \cite{Brown}
\begin{description}
\item[(N1)] If $N$ is a neighbourhood of $x$, then $x \in N$.
\item[(N2)] If $N$ is a subset of $X$ containing a neighbourhood of
$x$, then $N$ is a neighbourhood of $x$.
\item[(N3)] The intersection of two neighbourhoods of $x$ is again a
neighbourhood of $x$.
\item[(N4)] Any neighbourhood $N$ of $x$ contains a neighbourhood $M$
of $x$ such that $N$ is a neighbourhood of each point of $M$.
\end{description}
The pair $(X,\N)$ is then called a {\it topological
space}. Furthermore, a {\it base} for the neighbourhoods at $x$ is a
set $\Bas(x)$ of neighbourhoods of $x$ such that every neighbourhood
$N$ of $x$ contains an element $b \in \Bas(x)$. Now we define the path
$q(x)$ to be a neighbourhood base for $x$, and a subset $N \subset X$
to be a neighbourhood of $x$ if and only if there exists a $b \in
q(x)$ that is contained in $N$. The result is indeed a neighbourhood
topology on $X$:
\begin{theorem}[Trees and neighbourhood topology]
\label{TreeAndTopology}
Every tree structure $\B(X)$ over $X$ defines a neighbourhood topology
on $X$.
\end{theorem}
\begin{proof}
If $N$ is a neighbourhood of $x$ in our sense then it contains an
element $b \in q(x)$ and therefore contains $x$ as an element, even if
the path $q(x)$, or the tree $\B$, does not contain $\{x\}$ as an
element; thus, (N1) is satisfied. (N2) is fulfilled automatically by
our definition. Let $N$ and $N'$ be two neighbourhoods of $x$; then
they both contain elements $b$ and $b'$ of the same path $q(x)$, and
hence at least one of the relations $b' \supset b$ or $b \supset b'$
is satisfied. We can assume without loss of generality that the latter
is the case; then the intersection of $N$ and $N'$ contains $b'$ and
hence is a neighbourhood of $x$, thus (N3) is satisfied. Finally, let
$N$ be a neighbourhood of $x$; then $N$ contains some $b \in q(x)$,
which itself is a neighbourhood of $x$. Then, for every $y \in b$, $b$
lies in the path of $y$, as follows from
proposition~\ref{ThePathOfPoints}. Hence $b$ is a neighbourhood of
$y$. Consequently, $N$ is a neighbourhood for each $y \in b$. This
shows that (N4) is satisfied.
\end{proof}

It is clear that the set of all possible tree structures over the
given set $X$ may be constrained in many different ways, for example,
by imposing the conditions discussed in
section~\ref{TreesPreservingPartition}. On each constrained set of
trees, the tree function $E$ will take a minimum, which is an
entropy-like quantity, and will single out those trees on the
constrained set on which the minimum is actually taken. The associated
trees then define preferred topologies on the underlying set by means
of the construction given above. We see that this looks distinctly
like an action principle for topologies on the set $X$, the role of
the action being played by the tree function, the degrees of freedom
being expressed by the different trees over $X$, and the minimal value
of the action=tree function $E$ being associated with an entropy-like
quantity.

\section{Summary}

We have presented a comprehensive account of a new mathematical
structure, called tree structure, which arises in the formalisation of
the operational aspects of information gaining. It was shown that a
given set of tree structures can be endowed with a tree function whose
value is related to the maximal number of yes-no questions which are
necessary to identify a given node in the tree. The question of
minimality of the tree function on these sets of trees can be
posed. It was shown that, on unconstrained trees, the minimal value of
the tree function is related to the dyadic logarithm of the number of
elements in the base set; whilst, on constrained sets of trees, the
tree function takes minima whose functional form is similar to the
Shannon-Wiener information, or entropy, of a probability
distribution. We have presented three natural axioms governing tree
structures. It was subsequently demonstrated that these axioms can be
related to the axioms describing neighbourhood topologies on a given
set. As a consequence, every tree structure defines a neighbourhood
topology on a set. The minimisation of a tree function on a set of
tree structures over a base set then opens up the possibility to
obtain preferred neighbourhood topologies, namely those which are
related to minimal trees over the given base set. This phenomenon has
the distinct flavour of an action principle, distinguishing certain
preferred neighbourhood topologies by means of a minimal principle.


\section*{Acknowledgements}

Hanno Hammer acknowledges support from EPSRC grant 
\newline
No.~GR/86300/01.



\end{document}